\documentclass[12pt,manuscript]{emulateapj}

\shorttitle{$\rho$ State Physics in GRS 1915+105}
\shortauthors{Neilsen et al.}

\begin{document}

\title{The Physics of the `Heartbeat' State of GRS 1915+105}

\author{Joseph Neilsen\altaffilmark{1,2}, Ronald
  A. Remillard\altaffilmark{3}, Julia C. Lee\altaffilmark{1,2}}
\altaffiltext{1}{Astronomy Department, Harvard University, Cambridge,
  MA 02138; jneilsen@cfa.harvard.edu}
\altaffiltext{2}{Harvard-Smithsonian Center for Astrophysics,
  Cambridge, MA 02138}  
\altaffiltext{3}{MIT Kavli Institute for Astrophysics and Space
  Research, Cambridge, MA 02139}

\begin{abstract}
We present the first detailed phase-resolved spectral analysis of
a joint \textit{Chandra} High Energy Transmission Grating Spectrometer
and Rossi X-ray Timing Explorer observation of the $\rho$ variability
class in the microquasar GRS 1915+105. The $\rho$ cycle displays a
high-amplitude, double-peaked flare that recurs roughly every 50 s,
and is sometimes referred to as the ``heartbeat'' oscillation.
The spectral and timing properties of the oscillation are consistent
with the radiation pressure instability and the evolution of a local
Eddington limit in the inner disk. We exploit strong variations in the
X-ray continuum, iron emission lines, and the accretion disk wind to
probe the accretion geometry over nearly six orders of magnitude in
distance from the black hole. At small scales (1--$10~R_{\rm g}$), we
detect a burst of bremsstrahlung emission that appears to occur when
a portion of the inner accretion disk evaporates due to
radiation pressure. Jet activity, as inferred from the appearance of a
short X-ray hard state, seems to be limited to times near minimum
luminosity, with a duty cycle of $\sim10\%.$ On larger scales
(10$^{5}$--10$^{6}R_{\rm g})$ we use detailed photoionization
arguments to track the relationship between the fast X-ray variability
and the accretion disk wind. For the first time, we are able to show
that changes in the broadband X-ray spectrum produce changes in the
structure and density of the accretion disk wind on timescales as
short as 5 seconds. These results clearly establish a causal link
between the X-ray oscillations and the disk wind and therefore support
the existence of a disk-jet-wind connection. Furthermore, our analysis
shows that the mass loss rate in the wind may be sufficient to cause
long-term oscillations in the accretion rate, leading to state
transitions in GRS 1915+105.
\end{abstract}


\section{INTRODUCTION}
\label{sec:intro}
Accreting stellar-mass black holes are known to exhibit different
accretion `states,' which are usually defined in terms of their X-ray
spectral shape and variability (\citealt{RM06} and references
therein). Physically, these states are intimately related to the
fundamental parameters of the accretion flow, e.g.\ the accretion
rate, the accretion disk geometry, the radiative efficiency of the
disk, and the role of outflows in the form of winds and jets. The fact
that X-ray binaries (XRBs) undergo state transitions highlights the
dynamic nature of accretion onto black holes and allows for the
possibility of determining not only the fundamental parameters of
accretion but also the physics that controls them. 

\defcitealias{B00}{B00}
Of all the known Galactic black holes, GRS 1915+105 is undoubtedly the
most prolific source of state transitions. Discovered as a transient
by GRANAT in 1992 (\citeauthor{CastroTirado92}), it has remained in outburst for
the last 18 years and is typically one of the very brightest sources
in the X-ray sky. It is also one of the most variable: its X-ray
lightcurve consists of at least 14 different patterns of variability,
most of which are high amplitude and highly-structured (\citealt{B00},
hereafter B00; \citealt{K02,Hannikainen05}). It is believed that many of these
phenomenologically-described variability classes, which are labeled
with Greek letters \citepalias{B00}, are limit cycles of accretion and
ejection in an unstable disk \citep{B97b,M98,T04,FB04}.

The diverse timing and spectral properties of these variability
classes make the behavior of GRS 1915+105 particularly difficult to
track in physical detail. Even in the $\chi$ state, one of the simplest
variability classes (with a relatively low X-ray flux, no structured
variability, and a hard spectrum), GRS 1915+105 is never as faint or
as hard as the canonical `hard' state of black hole binaries
\citep{Belloni10b}. On the other hand, the canonical black hole `hard' state
is associated with jet production, and GRS 1915+105 has been shown to
produce a jet during essentially every spectrally hard interval longer
than 100 seconds \citep{K02}. This fact suggests that despite some
differences in the spectral shape, the physics of jet formation is
probably the same for all Galactic black holes \citep{FB04}.

\defcitealias{NL09}{Paper I}
It is well-established that the properties of jets are
highly correlated with the accretion state. But there is now a 
growing body of evidence suggesting that hot accretion disk winds are
equally influenced by the accretion state, in part but not completely
due to changes in ionizing flux (\citealt{L02,M06b,M08,NL09},
hereafter Paper I, \citealt{U10,Blum10}). In \citetalias{NL09}, we demonstrated
that the strength of the accretion disk wind in GRS 1915+105 is
anticorrelated with the \textit{fractional} hard X-ray flux, which is
therefore a useful diagnostic of both the accretion state and outflow
physics (for a preliminary discussion of these lines, see
\citealt{M08}). The anticorrelation holds over many variability
classes and there are indications that it may hold for other black
holes as well (XTE J1650-500 \& GX 339-4, \citealt{M04a}; GRO J1655-40, 
\citealt{M08}; H1743-322, \citealt{M06b,Blum10}).

If outflows from stellar-mass black holes depend on the accretion
state, then any rapid variability should have observable consequences
for those outflows. This is particularly true for GRS 1915+105, where
the X-ray spectrum and the accretion flow can change drastically in
seconds. For example, it has been shown conclusively that 30-minute
radio oscillations are `baby jets' produced by ejection events in
cycles like the $\beta$ state
\citep{Fender97,PF97,E98a,M98,Mirabel99}. Accretion 
disk winds have been directly observed to vary on similar (2--10 ks)
timescales (GRS 1915+105: \citealt{L02}; \citealt*{U09}; \citealt{U10}; Cir X-1:
\citealt{Schulz02}) or longer (1A 0535+262:
\citealt{ReynoldsM10}). Flux-dependent studies have implied faster
variations in both emission (1 s; \citealt{Miller05}) and absorption
lines (300 s; \citealt{M06b}). However, 
the physical consequences of rapid variability on these winds have yet
to be tracked in detail. A potential link between X-ray variability
and disk winds would be especially interesting given our recent
demonstration of a wind-jet interaction in GRS 1915+105 (\citetalias{NL09}).

In order to explore the relationship between jets, winds, and fast
variations in the accretion disk, we have undertaken a detailed
investigation of the $\rho$ variability class in GRS 1915+105. Known
affectionately as the `heartbeat' state because of the resemblance of
its X-ray lightcurve to an electrocardiogram, the $\rho$ state (see
Fig.\ \ref{fig:lc}) is a $\sim50$-s oscillation consisting of a slow
rise followed by a series of short bright bursts with
amplitudes of order $5\times 10^{38}$ ergs s$^{-1}$ and strong changes
in the X-ray spectral hardness (\citealt*{TCS97}, hereafter TCS97;
\citealt{Vilhu98,Paul98}). Theoretical models suggest that this state is a
manifestation of the Lightman-Eardley instability, a limit cycle in
the radiation-pressure dominated inner accretion disk
(\citealt{Lightman74,B97b}; \citealt*{J00,Nayakshin00}; \citealt{JC05}).  
\defcitealias{TCS97}{TCS97}
\defcitealias{N11a}{Paper II}

Previously, we analyzed the average spectrum of a joint 
\textit{RXTE}/\textit {Chandra} observation of GRS 1915+105 in the 
$\rho$ state (\citetalias{NL09}). We found the average \textit{RXTE} continuum
during this observation to be relatively soft, with $\sim79\%$ of the
3--18 keV X-ray luminosity ($L_{\rm X}\sim4.9\times10^{38}$ ergs
s$^{-1}$) emitted below 8.6 keV. In the time-averaged high-resolution
X-ray spectrum from the \textit{Chandra} High-Energy Transmission
Grating Spectrometer (HETGS; \citealt{C05}), we detected an 
Fe\,{\sc xxvi} Ly$\alpha$ absorption line from the accretion disk wind
with an equivalent width of $-7.2\pm1.7$ eV and a blueshift of
$\sim1400$ km s$^{-1}$.

We follow up in this paper by tracking the $\rho$-phase-resolve timing
and spectral variability of this same joint \textit{RXTE}/\textit
{Chandra} HETGS observation of GRS 1915+105. For the very first time,
we detect significant variations in absorption lines in phase-binned
spectra, allowing us to assess real physical changes on timescales of
seconds, well below the dynamical time in the wind. We explain the
origin and evolution of the accretion disk wind via analysis of the
X-ray spectral variability. Our results indicate that each bright
burst has a significant impact on the accretion dynamics from the
innermost to the outermost regions of the accretion disk. 

In Section \ref{sec:2} we describe our observations and data
reduction. In Section \ref{sec:timing}, we define the phase of the
$\rho$ cycle, explore variations in the recurrence time, and analyze
phase-resolved power spectra. We perform joint spectral analysis with
\textit{RXTE} and the \textit{Chandra}~HETGS in Section
\ref{sec:spectra}. We discuss our results on outflow formation and
disk instabilities in Section \ref{sec:discuss}, summarize our 
understanding of the oscillation in Section \ref{sec:physics}, and
conclude in Section \ref{sec:conc}. 

\section{OBSERVATIONS AND DATA REDUCTION}
\label{sec:2}
\subsection{\textit{Chandra}~Data}
\label{sec:chandraobs}
GRS 1915+105 was observed with the \textit{Chandra}~HETGS on 2001 May
23 (08:25:38 UT), for 30.16 ks. In order to mitigate pileup, the data
were taken in Continuous Clocking Mode; events were recorded in Graded
format to reduce the risk of telemetry saturation. 

\begin{figure}
\centerline{\includegraphics[width=0.49\textwidth]{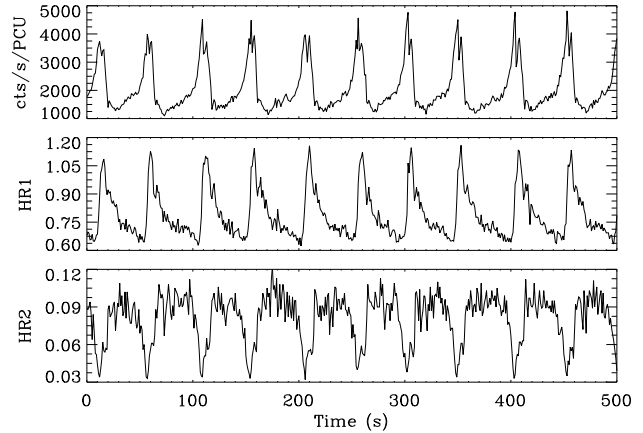}}
\caption{The 1-second PCA heartbeat lightcurve (top panel), HR1
  (middle panel), and HR2 (bottom panel). HR1$\equiv B/A$ and
  HR2$\equiv C/B$, where $A,~B,$ and $C$ are the PCA count rates in
  the 2.0--4.5 keV, 4.5--9.0 keV, and 9.0--30 keV bands. The heartbeat
  is composed of a slow rise or a shoulder followed by a double-peaked
  pulse. In HR1 the behavior is similar, although the shoulder follows
  the maximum rather than preceding it.} 
\label{fig:lc}
\end{figure}
We reduced and barycenter-corrected the \textit{Chandra}~data using
standard tools from the {\sc ciao} analysis suite, version 4.0. After 
reprocessing and filtering, we extracted High-Energy Grating
(HEG) spectra and created grating responses. We used the order-sorting
routine to remove the ACIS S4 readout streak, since the
\textit{destreak} tool can introduce spectral artifacts for bright
continuum sources like GRS 1915+105. We extracted 1-second lightcurves
with \textit{dmextract}.
\begin{figure*}
\centerline{\includegraphics[width=\textwidth]{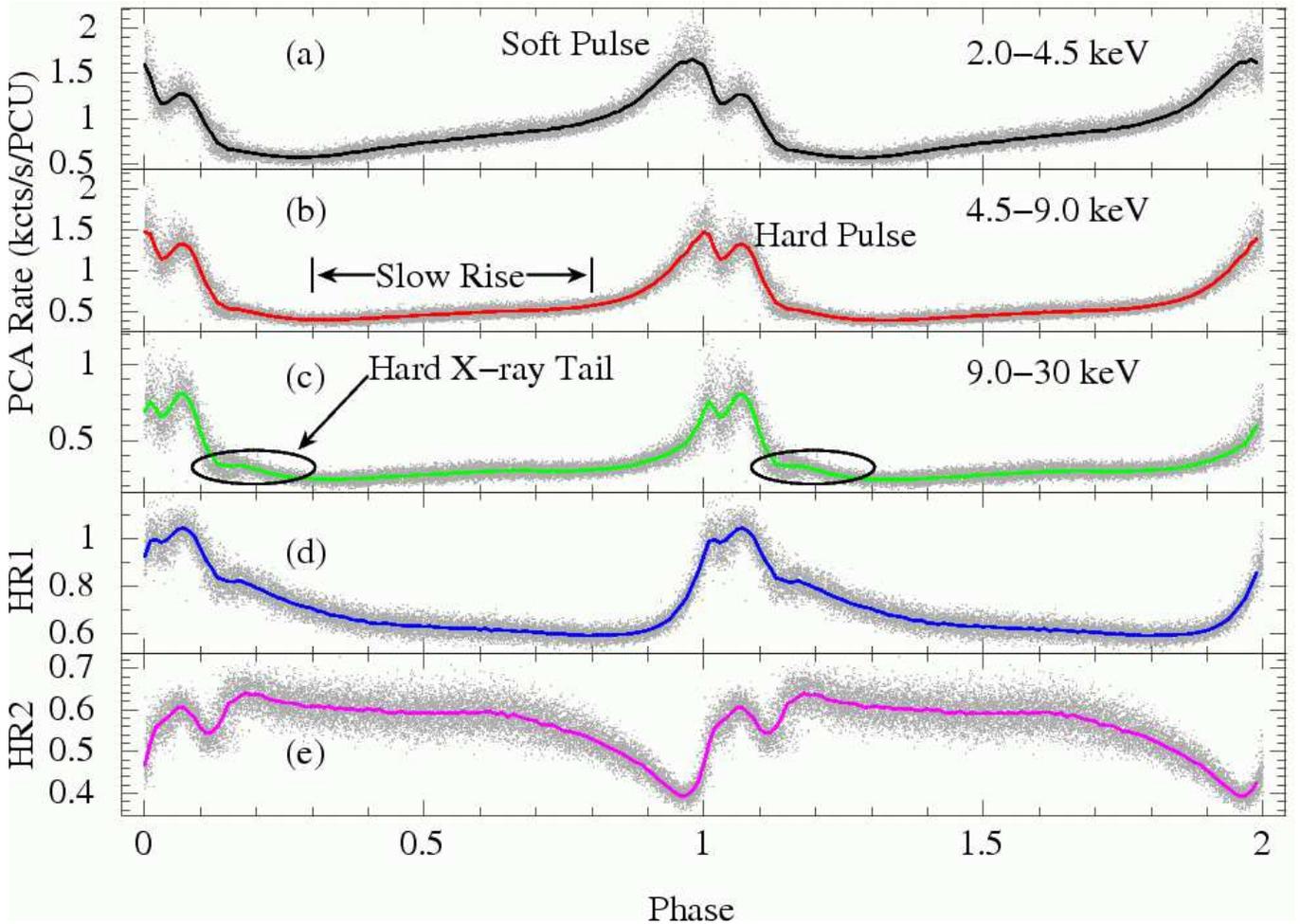}}
\caption{The phase-folded PCA heartbeat lightcurve in the (a)
  $A\equiv2$--4.5 keV, (b) $B\equiv4$.5--9.0 keV, and (c)
  $C\equiv9.0$--30 keV bands, along with (d) HR1 and (e) HR2, which
  are defined in Section \ref{sec:timing}. As discussed in
  the text, the mean period is about 50 seconds, so each major tick
  represents about 5 seconds. We have labeled the slow rise ($\sim25$
  s), the soft and hard pulses, and the hard X-ray tail ($\sim5-10$
  s). Two cycles are shown for clarity. The gray points in the
  background are the individual data points, showing remarkable
  consistency despite noticeable variations in the cycle period.} 
\label{fig:philc}
\end{figure*}

Due to incomplete calibration of Charge-Transfer Inefficiency (CTI) in
CC/Graded mode, there is some wavelength-dependent disagreement in the
continuum flux between spectral orders of the HEG (and the MEG, which
we do not consider here because of its lower spectral
resolution). For this reason, it is not currently possible to fit a
physical continuum model to the HETGS data. Instead, we fit the
individual spectra with polynomials to model the local continuum and
use Gaussians for line features found in the combined residuals.\\ 

\subsection{\textit{RXTE} Data}
\label{sec:rxteobs}
During the \textit{Chandra}~observation, \textit{RXTE}
made a pointed observation of GRS 1915+105, beginning on 2001 May 23
at 11:08:20.8 UT and lasting 21.9 ks (elapsed) with 13.7 ks exposure
time. We select all available data subject to the following
constraints: (1) the Earth-limb elevation angle is above 3$^\circ$;
(2) the spacecraft is outside the South Atlantic Anomaly; (3) the
offset angle from GRS 1915+105 is less than 0.02$^\circ$. In this
paper, we analyze the data from the Proportional Counter Array (PCA),
which covers the 2--60 keV band.

For timing analysis, we make use of the data from the binned mode
B\_8ms\_16A\_0\_35\_H\_4P, which covers the 2.0--14.8 keV band at 7.8 
ms time resolution, and the event mode E\_16us\_16B\_36\_1s, which
covers the 14.8--60 keV band at 15.3 $\mu s$ time resolution. We
extract 1-second barycentered, dead-time-corrected, and background
subtracted lightcurves from each of these modes for timing analysis,
and use our own software to create power spectra, subtracting the
dead-time-corrected Poisson noise level after \citeauthor*{MRG97} (1997,
hereafter MRG97). For high-S/N hardness ratios, we use the energy
bands 2--4.5 keV, 4.5--9.0 keV, and 9.0--30 keV, which have roughly
equal count rates. 
\defcitealias{MRG97}{MRG97}

For spectral  analysis, the strong fast variability of the $\rho$
state precludes use of the Standard-2 129-channel spectra, since the
16-second time resolution of the Standard-2 data is too coarse for the
50-second cycles. To compensate, we use the binned mode and event mode
data to create 32-channel spectra at 1-s time resolution (see,
e.g.\ \citetalias{TCS97}, \citealt{B97b}). We treat intervals with different
combinations of PCUs separately. Our observation features the PCUs
operating in combinations \{0, 2, 3, 4\} and \{1, 2, 3, 4\}, with
roughly equal exposure times.

\section{TIMING ANALYSIS}
\label{sec:timing}
We show a representative portion of the PCA 1-second lightcurve in
Figure \ref{fig:lc}, along with PCA hardness ratios HR1 and
HR2 (here defined as the ratio B/A and C/B, where A, B, and C are the
count rates in the 2.0--4.5 keV, 4.5--9 keV, and 9--30 keV bands,
respectively). The lightcurve consists of a $\sim25$ second slow rise,
followed by a short burst, which is typically double-peaked (see also
Figure \ref{fig:philc}). HR1 looks rather like the mirror image of the
count rate (i.e.\ flipped horizontally about its maximum), except that
HR1 is typically single-peaked. Other instances of the $\rho$ state
have shown different behavior, as the number and relative strengths of
the peaks can vary (\citealt{Massaro10}, Neilsen et al.\ 2011, in
preparation).

It is our main goal in this section to characterize the timing
properties of the $\rho$ state by tracking the arrival times of each
individual burst. We define a phase ephemeris for the oscillation with
$\phi\equiv0$ at the time of maximum count rate (see the Appendix for
the motivation for this choice and the details of our method). This
timing analysis is the foundation of all of the results presented in
this paper, as it allows us to study the evolution of the X-ray
lightcurve, power spectrum, broadband X-ray spectrum, and the
high-resolution X-ray spectrum as a function of phase (instead of
time).

\subsection{Phase Ephemeris and Folded Lightcurve}
\label{sec:philc}
\citetalias{MRG97} tracked individual QPO waves in GRS 1915+105 by fitting
functional templates (e.g.\ sinusoids or Gaussians) to the PCA
lightcurve. Given the high amplitude and unusual shape of the
lightcurve described here, we opt for a modified version of the
\citetalias{MRG97} technique (see the Appendix for more details). 

First, we take a single representative cycle from the data and
cross-correlate it with the entire lightcurve. Maxima in the
normalized cross-correlation values then indicate the times when
$\phi=0$ (peaks in the count rate). We then fold the data on this
first set of $\phi=0$ times to obtain the average folded
lightcurve. The process is iterated, with the folded lightcurve serving
as a new template, to obtain the final set of $\phi=0$ times. This
analysis results in 623 peak times from the \textit{Chandra}
lightcurve and 273 peak times from the PCA lightcurve, which should be
accurate to 0.1 seconds or better for \textit{RXTE} and 0.3 seconds or
better for \textit{Chandra}. The \textit{RXTE} uncertainty is
estimated from the variations produced by the use of different
cross-correlation templates, while the \textit{Chandra} uncertainty is
derived from the offsets in $\phi=0$ times for cycles co-measured with
\textit{RXTE} (see the Appendix for details). 

We also explored the 3-D pathway of the $\rho$ cycle
through the hardness-hardness-intensity diagram (HHID;
see \citealt{Soleri08} for an example of this method) to consider
using the mean pathway to construct a set of phase-resolved 
spectra that could be analyzed and compared to the results derived
from the cross-correlation ephemeris method. We find that the HHID
method is less suitable for the $\rho$ cycle because the statistical 
noise for 1 s bins causes significant scatter in the HHID
(particularly in the hard color dimension), and because the rho cycle 
shows a tight loop in all three parameters over the phase interval
0.04-0.12, causing some degeneracy for phase tracking based on the
HHID. Thus we believe cross-correlations provide the optimal
characterization of the cyclic variability. With a phase ephemeris
defined, we can create phase-folded lightcurves and hardness ratios
(Figure \ref{fig:philc}). It is clear from this figure that
phase-folding is a very effective way to characterize the oscillation,
since the individual cycles have very similar shapes: the
cycle-to-cycle variability is always less than 20\% for a given phase.

To help diagnose whether there is systematic noise introduced by using
the broadband count rate to track the heartbeat cycles, compared to a
parameter related to spectral shape, we conducted the following
test. We used the PCA soft color, i.e.\ the ratio of source counts at
6-12 keV versus 2-6 keV, as an alternative quantity to compute cross 
correlations to define the times when each heartbeat cycle begins.
The procedures were otherwise left intact. We note that the soft
color usually shows one maximum per cycle, which is offset from the
times of maximum count rate. The comparison of 239 cycle start times
derived from these two tracking parameters (\textit{RXTE} data; mean
cycle period 50.488 s) yields an average offset of 3.274 s, with a sample
standard deviation of 0.283 s. It is the latter quantity that measures
systematic noise differences between the method that uses the count
rate to track cycles, versus soft X-ray color. We conclude
that neither our phase-resolved lightcurves and binned spectra nor our
analysis conclusions would be significantly changed if the heartbeat
phases were defined by soft color, since the phase deviations (0.0056)
are much less than the spectral bin size (0.02). 
\begin{figure}
\centerline{\includegraphics[angle=270,width=0.49\textwidth]{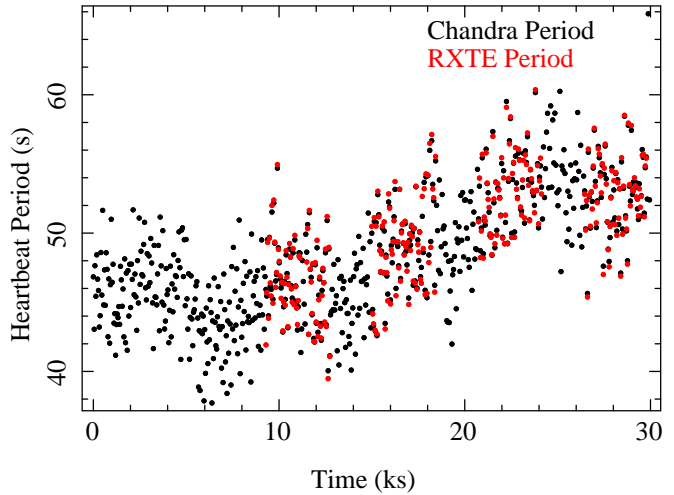}}
\caption{The heartbeat period during our observations, measured as the
  time difference between successive cycles. The periods measured by
  \textit{Chandra} (black) and \textit{RXTE} (red) show very good
  agreement. Interestingly, the period exhibits considerable
  variability on short timescales and drifts slowly over long
  timescales.\vspace{3mm}}
\label{fig:heartbeatperiod}
\end{figure}

During our observation, each folded heartbeat is a slow
rise $(\phi=0.3-0.8 \approx25$ s), followed by a double-peaked pulse
($\phi=0.8-0.1\approx15$ s) and a hard X-ray tail lasting 5--10
seconds (see $\phi\sim0.2$ in panel c). Generally speaking, the slow
rise is spectrally hard, the first pulse is soft, and the second pulse
is moderately hard, but it is obvious even from Figure \ref{fig:philc}
that such statements must be made and interpreted only loosely, since
HR1 and HR2 are more or less anticorrelated.

Other authors \citep{JC05,Massaro10} have already demonstrated that there is
typically a delay between the soft and hard X-ray pulses, as well
as a strong relationship between the times between these two bursts
and the cycle period. Because HR1 peaks at the hard pulse, we see
that the hard X-ray pulse lags the soft X-ray pulse by an average of
3.3 seconds in our observation. This lag increases with the length of
the cycle and appears to be a constant 6.5\% of the cycle
duration. This surprising result implies that the delay between soft
and hard X-ray pulses scales with the same clock as the entire
oscillation. It also implies that using an X-ray color to define
$\phi\equiv0$ simply introduces a phase shift into our results. Note
that the variations in the lag are 
not large enough to interfere with our cross-correlation or
phase-folding. We will return to the delay in Sections \ref{sec:pca}
and \ref{sec:eject}, where we analyze the X-ray continuum.

\subsection{The Cycle Period}
\label{sec:period}
We define the cycle period as a discrete quantity given by the time
difference between two successive times of $\phi=0$ in the
lightcurve. We show the heartbeat period for \textit{Chandra} and
\textit{RXTE} in Figure \ref{fig:heartbeatperiod} (black and red,
respectively). Two points are immediately clear from this figure, both
of which imply that the oscillation period in the $\rho$ state is
similar to that of low-frequency QPOs observed in GRS 1915+105: 
\begin{enumerate} 
\item The heartbeat cycle period exhibits a slow secular drift from
  $\sim45$ seconds to $\sim53$ seconds over 30 ks.
\item There is a substantial amount of scatter with an amplitude
  much higher than the secular drift. This scatter is also much larger
  than the uncertainty in the cycle period, 
which is of order 0.1 seconds for \textit{RXTE} measurements
(see Section \ref{sec:philc}).
\end{enumerate}

The scatter in the period exemplifies the quasi-periodic nature of the
heartbeat cycle: consecutive periods may vary by 5--10 seconds or
more. The variability in Figure \ref{fig:heartbeatperiod} is
reminiscent of the behavior of QPOs in a variety of X-ray states of
GRS 1915+105 as seen by \citetalias{MRG97}, who tracked the arrival times of
individual oscillations for several QPOs with frequencies from 0.067
Hz to 1.8 Hz.

MGR97 were able to demonstrate that the arrival times of these QPOs
were well-described by uncorrelated Gaussian noise. For the $\rho$
state, we find essentially no correlation between successive periods
(the autocorrelation coefficient is $r=-0.018),$ and the cumulative
distribution of periods is fit well with a two-Gaussian model. These
Gaussians have mean periods $\mu_1=46.0$ s and $\mu_2=52.8$ s, with
standard deviations $\sigma_1=2.2$ s and $\sigma_2=2.1$ s. 

The fractional scatter in the period measured here is comparable to 
the $\mu=14.93$ s and $\sigma=0.72$ s found by \citetalias{MRG97} for
the 0.067 Hz 
QPO, and is actually less than the scatter measured for the other
QPOs they tracked. On the other hand, the drift in the $\rho$ period
is much larger (roughly 40 phase units per 1000 cycles compared to 13 
phase units per 1000 cycles for the 0.65 Hz QPO). Given its strong
spectral variations, the $\rho$ cycle is much more complex than a
normal QPO, but the variability in its period is quantitatively
similar. Given this similarity, it is interesting that a QPO at 6--15
Hz is present in the power spectrum during much of the $\rho$ state,
which is the subject of Section \ref{sec:powspec}.

\subsection{Power Spectra}
\label{sec:powspec}
The phase-folded HR2 curve (Fig. 2) shows two maxima that indicate
enhanced hard X-ray components during the hard pulse
($\phi=0.05-0.1$) and the hard X-ray tail (beginning near
$\phi=0.15$). To help diagnose the spectral conditions during these
intervals, we compute power density spectra (PDS) for each second of
source exposure with the RXTE PCA, using the count rate in the 2--37.9
keV band.  In addition to subtracting the deadtime-corrected Poisson
noise (see Section 2.2), we normalize the PDS to units of $(rms /
mean)^2$ Hz$^{-1}$ (\citetalias{MRG97}).

To search for possible contributions from a jet during the
heartbeat cycle, we hope to exploit the body of evidence that
associates a steady radio jet with the properties of the X-ray hard
state. The hard state has the signatures of enhanced power ($rms >
0.1$) and a power continuum with ``band-limited'' shape, i.e., power
density that appears flat at frequencies below a few Hz
\citep{Muno01,Fender06,RM06}. We note that measurements of $rms$ in the
literature often integrate the power over the range 0.1-10 Hz, but in
this study a lower limit of 2 Hz is dictated by the Nyquist frequency
for data samples taken at every second.

\begin{figure}
\centerline{\includegraphics[width=0.49\textwidth]{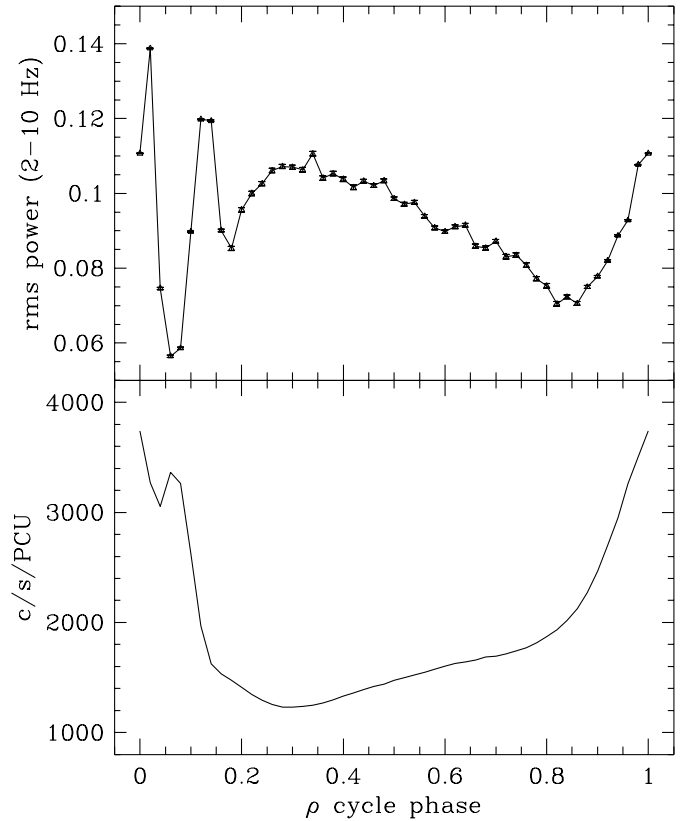}}
\caption{2--10 Hz $rms$ power as a function of cycle phase during the
  $\rho$ state. We find $rms>0.1$ (one of the criteria for the
  jet-producing X-ray hard state) in three phase windows:
  $\phi=0.02,~\phi=0.12-0.14,$ and in a broad interval near
  $\phi=0.3.$ While none of these windows corresponds to the hard
  pulse in the X-ray lightcurve, the broad interval does overlap with
  the hard X-ray tail. Thus the later stages of the hard X-ray tail
  may correspond to a short hard state.}
\label{fig:rms}
\end{figure}

\begin{figure}
\centerline{\includegraphics[width=0.49\textwidth]{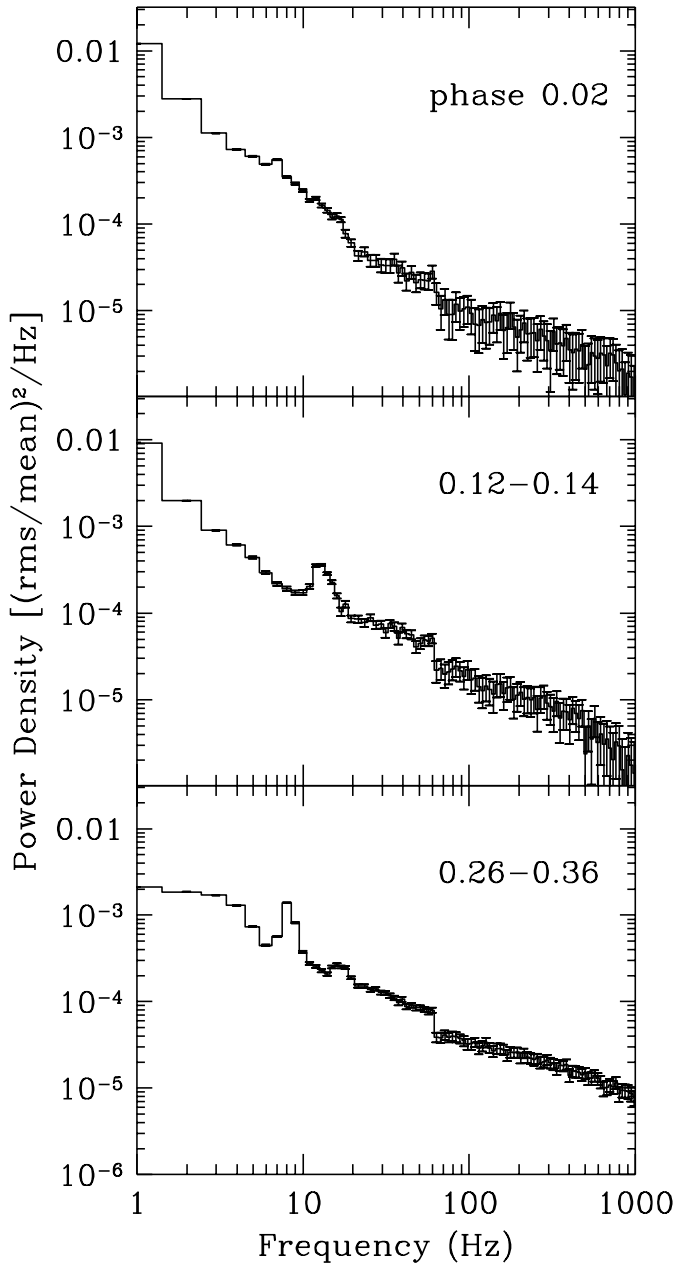}}
\caption{Power density spectra (PDS) corresponding to the three $rms$
  maxima in Figure \ref{fig:rms}: from top to bottom, $\phi=0.02,
  ~\phi=0.12-0.14,$ and $\phi\sim0.26-0.36$. During the first and
  second maxima, the PDS has a power-law shape. It is essentially
  featureless in the first maximum, but a QPO appears at 13 Hz in the 
  second maximum. The third maximum shows band-limited noise that is
  consistent with the hard state, along with a QPO at $\sim8$ Hz.} 
\label{fig:powspec}
\end{figure}

We first examine the phase-folded values of $rms$ power.  This is
derived by integrating each PDS over the range 2--10 Hz, and then
averaging the results in each of 50 phase bins, using the heartbeat
cycle phase ephemeris described in Section 3.1.  The phase-folded
$rms$ curve is shown in Fig.~\ref{fig:rms}. Here we see three maxima
that exceed values of 0.1. These occur in narrow phase windows at 0.02
and 0.12-0.14, and in a broad interval near $\phi= 0.3$. The first two
maxima are likely associated with the intrinsic variability of the
oscillation, particularly the sharp decay of the soft and hard
pulses, but we cannot rule out contributions from other
sources of normal variability. However, none of the three intervals
coincides with the hard pulse itself, which is instead aligned with a
dip (to $rms<0.06$) in Fig.~\ref{fig:rms}. On the other hand, there is
overlap between the later portions of the hard tail and the broad
interval of enhanced $rms$ after $\phi=0.25$.

The PDS during the phases corresponding with the three $rms$ maxima
are shown in Fig.\ \ref{fig:powspec}.  Although the intrinsic
variability of the $\rho$ cycle may artificially enhance the rms
variability during the first two intervals and could bias the
overall slope of the PDS, it is very unlikely to affect the
qualitative shape of power continuum. At phase 0.02, the power density 
spectrum is featureless and the power density ($P_{\nu}$) declines
with frequency ($\nu$) at a rate slightly steeper than $P_{\nu}
\propto\nu^{-1}$. This PDS does not resemble the X-ray hard state. At
the time of the second $rms$ spike (phase 0.12-0.14), the power
continuum again appears with a power-law shape, but there is
additionally a QPO near 13 Hz. Finally, the PDS during the third phase
interval with enhanced rms power does show a band-limited shape that
resembles the hard state. Looking at the individual phase-binned PDSs,
we find that similar shapes in the power continuum occur during the
phase range 0.20-0.36, indicating the most likely interval for
contributions from a jet that may temporarily form during each
heartbeat cycle.

The low-frequency QPO is detected during phase intervals 0.12-0.82 and
0.96-0.00, 
and it is therefore much more prevalent than either an elevated $rms$ 
or a power continuum shape that resembles the hard state. The QPO
frequency is 14 Hz at $\phi=0.12$, after which it slowly decreases to
7.3 Hz near $\phi=0.52$, and ends near 8 Hz at $\phi=0.82$. As
suggested by the bottom panel of Figure \ref{fig:powspec}, a first
harmonic is usually present. The presence of a QPO while
the source evolves through a long, hard dip in the X-ray lightcurve is
reminiscent of the $\beta$ cycle in GRS 1915+105, which is a 30-min
cycle tied to the formation of impulsive radio jets
(\citealt*{Markwardt99,Mikles06}; \citealt{M98,FB04}). Given that the
X-ray spectra during the 
$\beta$ cycle were successfully interpreted as the combination of
thermal radiation from an accretion disk plus a hard X-ray power law
attributed to inverse Compton emission, the QPO behavior in the long
hard dip of the $\rho$ cycle may justify application of the inverse
Compton model to this state. 

However, this leaves behind two puzzles. First, the hard pulse 
coincides with a phase interval that exhibits neither high $rms$
values nor a low-frequency X-ray QPO, and so we find no temporal
signatures that might promote spectral interpretation of this hard
X-ray component via Comptonization. Secondly, there is another brief
appearance of a QPO at phases 0.96-0.00 (with frequency that shifts  
from 6 to 7 Hz). This QPO is weaker than the 7--14 Hz feature and
lacks a first harmonic, so it is difficult to link this short interval
to the longer QPO episode. \citet*{Soleri08} also detected transient
QPOs in GRS 1915+105, associating them with state transitions. But
because our transient QPO does not coincide with an obvious state
transition, the relation between these features is unclear.

Finally, in some phase intervals ($0.02-0.06,~0.46-0.50$, and
$0.68-0.74$) there is also weak evidence ($2.8 - 3.1 \sigma$) of a QPO
($rms\sim 1$\%) near 60 Hz. This is of interest because GRS 1915+105
is known to exhibit a 67 Hz QPO \citepalias{MRG97} that can be
particularly strong in $\gamma$-type lightcurves. A hint of this
feature is seen at 59 Hz in the top panel of Fig.\ \ref{fig:powspec}. 
At other phases the feature looks more like an edge, as
suggested by the bottom panel of Fig.\ \ref{fig:powspec}. Examination
of the PDS in individual phase bins suggests that this edge is
\textit{not} an artifact of a stronger QPO moving to lower frequency.
These features are too weak to support further analysis, and
so this topic is left to a future generation of instruments with
larger collecting area.

\section{SPECTRAL VARIABILITY ANALYSIS}
\label{sec:spectra}
In the following subsections, we explore in detail the spectral
variability of the $\rho$ state, using \textit{RXTE} to study
the broadband X-ray properties of the oscillation and the
\textit{Chandra} HETGS to probe the known accretion disk wind as a
function of $\rho$ cycle phase. In the previous section, we
demonstrated fast spectral evolution in the $\rho$ state (see also
\citetalias{TCS97}; \citetalias{B00}), but it remains to be seen if and how
accretion disk winds and jets participate in this variability. Here we
will show that the changes in the X-ray continuum are related to
significant variations in the accretion disk wind, both occurring on
timescales $\lesssim5$ seconds. This constitutes the very first probe
of disk wind physics on such short timescales. All spectral fitting is
done in ISIS \citep{HD00,Houck02}. We assume a distance and inclination of
$D=11.2$ kpc and $i=66^{\circ}$ \citep{F99}; we fix $N_{\rm H}=5\times
10^{22}$ cm$^{-2}$ (\citealt{L02} and references therein). 

\begin{figure}
\centerline{\includegraphics[width=0.49\textwidth]{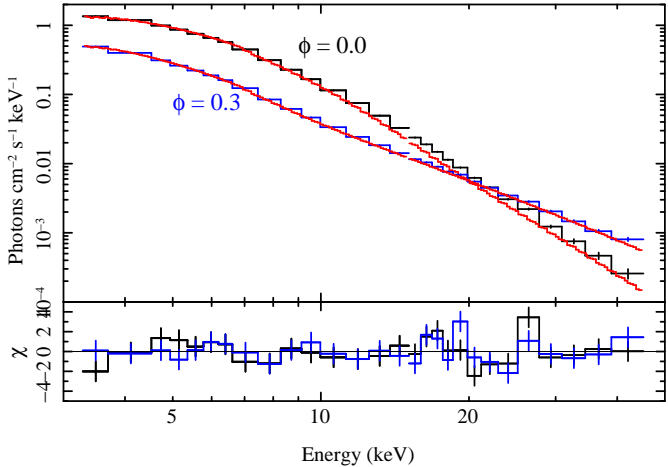}}
\caption{PCA spectra and residuals for two different phases of the
  heartbeat cycle. The fainter spectrum from the cycle minimum (phase
  $\phi=0.3$) is significantly flatter than the brighter spectrum at
  the peak, $\phi=0.$ See Section \ref{sec:pca} for details of our
  continuum models.}
\label{fig:pca_spec}
\end{figure}

\subsection{\textit{RXTE} PCA}
\label{sec:pca}
For our variability analysis, we apply our derived phase ephemeris
from Section \ref{sec:philc} to extract phase-resolved spectra: for
each 1-second PCA spectrum, we compute the $\rho$ cycle phase and
average the results using 50 phase bins. At each phase, we fit the 
spectrum from 3.3 to 45 keV. X-ray spectra of black hole binaries have
been modeled as the combination of various soft and hard components,
which may include but are not limited to: thermal radiation from the 
accretion disk, power law emission, an explicit treatment of
Comptonization, or bremsstrahlung emission. For the highly 
variable states of GRS 1915+105, a model consisting of a
multi-temperature accretion disk plus some type of power law has been
especially effective at tracking the relative changes in the accretion
disk and Comptonization (see e.g.\ \citealt{B97b,Migliari03}).
 
We focus here on the variations of the accretion disk and hard
X-ray components inferred from several different continuum models. We
show two example phase-resolved PCA spectra (from the peak and minimum
of the oscillation) in Figure \ref{fig:pca_spec}. As expected from our  
energy-resolved lightcurves, the spectrum at the peak of the count
rate ($\phi=0$) is significantly brighter and softer than the flat
spectrum at the cycle minimum ($\phi=0.3$).  

\subsubsection{Model 1: {\tt simpl}}
\label{sec:model1}
Our initial model for the X-ray continuum consists of five components:
cold absorption ({\tt tbabs}; \citealt*{Wilms00}), a hot disk component
({\tt ezdiskbb}; \citealt{Zimmerman05}), an emission line at 6.4 keV ({\tt
  egauss}), a high-energy cutoff ({\tt highecut}, whose formula is
given by $\exp(E/E_{\rm fold})$), plus a scattering component ({\tt simpl}; 
\citealt{Steiner09a}). {\tt simpl} is a convolution model that takes
any seed spectrum and scatters a fraction $f_{\rm SC}$ of the photons
into a power law. The high-energy cutoff is required to account for
curvature in the hard X-ray spectrum, and we choose {\tt simpl}
because it conserves photons and avoids the divergence of the {\tt
  powerlaw} model from realistic expectations of Comptonization at low
energies. We allow the different PCU combinations to have different 
parameters, although we tie the high-energy cutoff parameters together
for better constraints. In the end, our best fit parameters for
the different PCU sets typically differ by $\lesssim1\sigma$. Model 1
provides an excellent statistical description of the X-ray continuum
at nearly all phases of the $\rho$ cycle (overall
$\chi^{2}/\nu=2301.9/2150=1.07$), but the parameters during 
the hard pulse ($\phi\sim0.06$) are somewhat puzzling. See below and  
Section \ref{sec:bremss} for more details and alternative
explanations.

\begin{figure*}[t]
\centerline{\includegraphics[width=\textwidth]{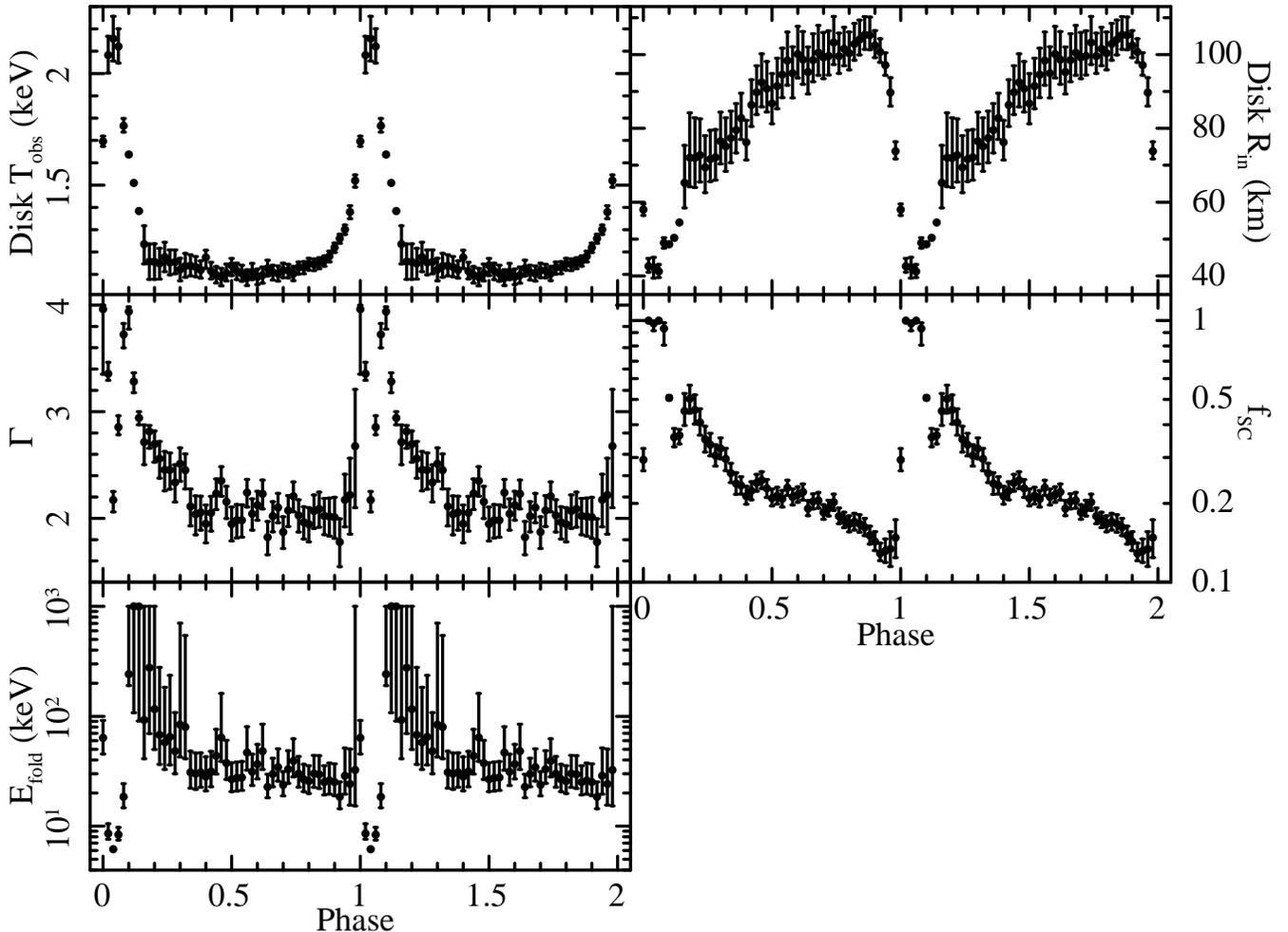}}
\caption{Model 1 fit parameters and $1\sigma$ error bars as a
  function of phase for the PCA spectra (averaged over different PCU
  combinations). Two cycles are shown for clarity; since the average
  period is 50.33 seconds, each phase bin corresponds to
  roughly 1 second of real-time variability. The
  temperature and radius of the accretion disk vary significantly
  throughout the cycle, reminiscent of changes in the $\kappa/\lambda$
  states analyzed by \citet{B97a,B97b}. Because {\tt simpl} counts
  disk photons to create the power law ($f_{\rm SC}$), Model 1
  uniquely tracks the disk parameters \textit{prior} to
  Comptonization, i.e.\ scaling up the disk normalization by a factor 
  $1-f_{\rm SC}$. See \ref{sec:model1} for details of the model
  fits.\vspace{3mm}}
\label{fig:model1}
\end{figure*}

We show the results of our broadband spectral fits in Figure
\ref{fig:model1}. In these fits, the observed maximum temperature in
the disk $T_{\rm obs}$ hovers around 1.1--1.2 keV for the majority
of the cycle, spiking sharply to $\sim2.2$ keV during the X-ray
burst. At the same time, the inner radius of the disk is relatively
large (70--110 km) and grows during the slow rise, dropping swiftly to
$\sim40$ km at the end of the soft pulse. The inner radius 
$R_{\rm in}$ is related to the {\tt ezdiskbb} normalization 
$N_{\tt disk}$ as \citep{Zimmerman05}:
\begin{equation}\label{eq:rin}
N_{\tt disk}=\frac{1}{f^4}\left(\frac{R_{\rm in}}{D}\right)^2\cos~i.
\end{equation} Because {\tt simpl} counts photons when
calculating the disk normalization and the power law ($f_{\rm SC}$),
$N_{\rm disk}$ in Model 1 uniquely tracks the disk parameters
\textit{prior} to Comptonization. For the high luminosities here, we
use a color correction factor $f=1.9$ (J.\ Steiner, private
communication). The results are qualitatively similar to the fast
variations seen by \citet{B97a,B97b} in the $\kappa$ and $\lambda$
states, although their disk radii are a factor of $\sim2$ lower than
ours. This difference appears to be mainly due to our use of {\tt
  ezdiskbb} rather than {\tt diskbb}. Thus it does not indicate a
physical difference between the $\rho$ state and the $\kappa$ and
$\lambda$ states.

During the slow rise the photon index $\Gamma$ is stable
around 2.1, which is relatively steep for a hard state but not unusual
for GRS 1915+105. But near $\phi=0$, the $\Gamma$ displays a
double-peaked phase dependence, rising quickly to its
upper limit ($\Gamma=4$; \citealt{Steiner09a}), dropping sharply back to
$\sim2.5$ during the hard pulse, then peaking again near $4$. After
the pulses, $\Gamma$ decays exponentially back to 2 with an
$e$-folding interval $\Delta\phi\sim0.1$. 

Meanwhile, the scattering fraction $f_{\rm SC}$ decreases slowly
during the slow rise and reaches a minimum $\sim0.13$ near
$\phi=0.95$. Then, during the hard pulse at $\phi\sim0.06$, $f_{\rm
  SC}$ spikes sharply to 1; there is a second, broader peak with
$f_{\rm SC}\sim0.6$ during the hard X-ray tail
(Fig. \ref{fig:model1}). We interpret this behavior as an indication
that during the hard pulse, the X-ray spectrum is completely dominated
by scattering and there is essentially no direct disk
component. Still, the disk dominates the light during the soft pulse,
contributing as much as 70\% of the observed flux just before
$\phi=0,$ and roughly 25\% during the hard X-ray tail. 

The high-energy cutoff exhibits similarly strong variations around the
phase of the hard pulse. During most of the cycle, $E_{\rm fold}$ is
steady around 30 keV, although the spectrum of the hard X-ray tail is
effectively consistent with no cutoff, or with $E_{\rm fold}\lesssim1$
MeV. But during the hard pulse, $E_{\rm fold}$ drops sharply to 6
keV. Thus Model 1 leads to the interpretation of the hard pulse as a
period of very strong (i.e.\ Compton thick) scattering by relatively
cool electrons. 

The flux ($F_{\rm 6.4~keV}$) in the Gaussian emission line (see
Section \ref{sec:feline}) declines very slowly for most of the cycle, 
but peaks during the soft pulse. With the exception of this short
pulse, the line flux is very strongly correlated with the line width
$\sigma_{\rm 6.4~keV},$ which varies between about 0.5 keV and 1.5
keV. The correlation coefficient for these two parameters is
$r=0.88$. At the \textit{RXTE} spectral resolution, it is difficult to
reliably decouple the variations in the line width and the line flux,
so the true variability of the feature is unclear. The line is too
broad to be observable in the \textit{Chandra} HETGS
spectrum. However, if we include only the phases of the cycle away
from the pulses, $\phi=0.15-0.8,$ we find a moderate correlation
between the line flux and the {\tt simpl} scattering fraction
($r\sim0.63$). We will return to this in Section \ref{sec:feline}.
 
\begin{figure*}
\centerline{\includegraphics[width=\textwidth]{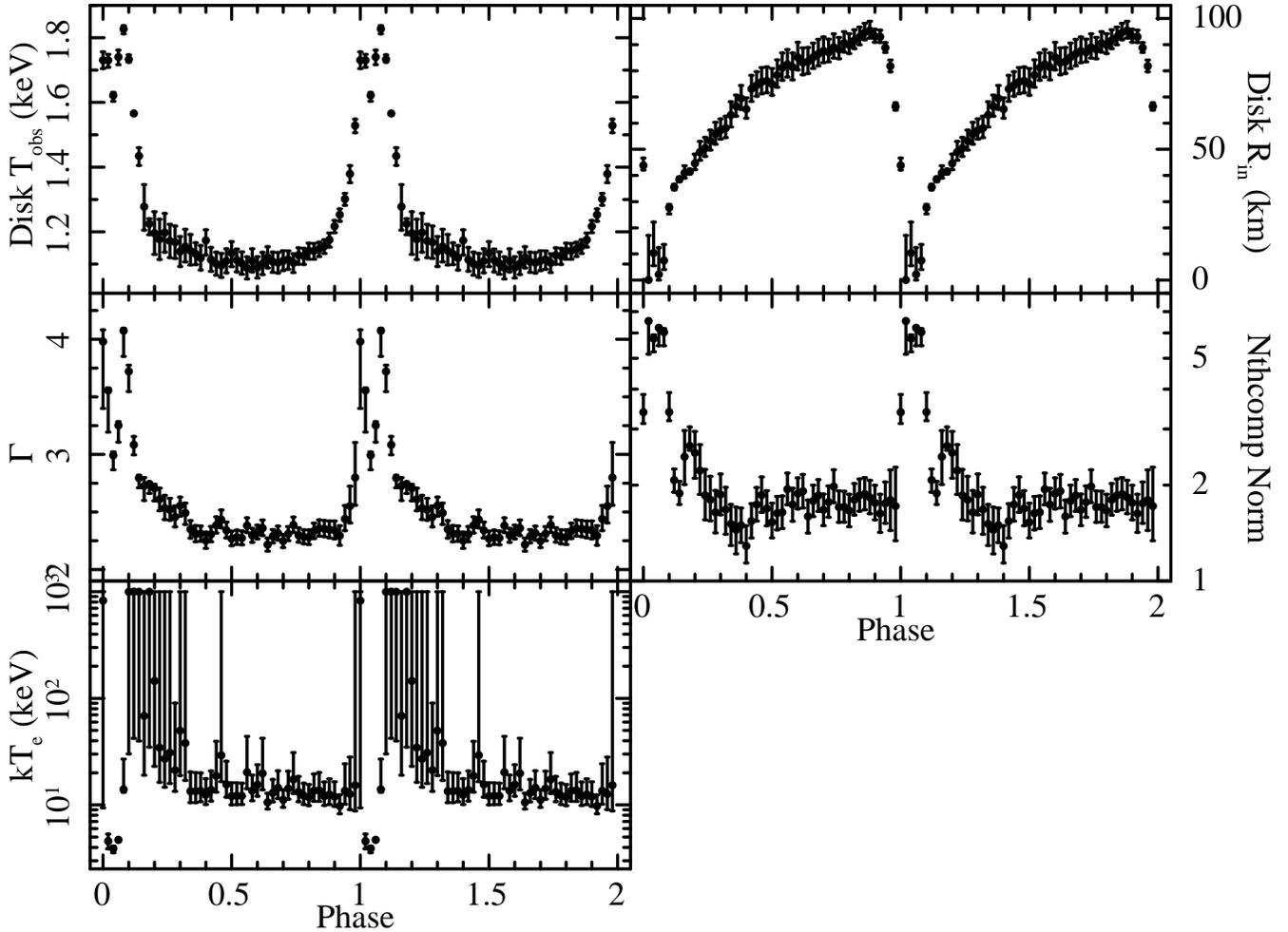}}
\caption{Model 2 fit parameters and $1\sigma$ error bars as a
  function of phase for the PCA spectra (averaged over different PCU
  combinations). Two cycles are shown for clarity; since the average
  period is 50.33 seconds, each phase bin corresponds to
  roughly 1 second of real-time variability. The disk
  temperature and radius are very similar to those of Model 1. The
  {\tt nthcomp} seed photon temperature is tied to the accretion disk
  temperature, so that both parameters are well-constrained even when
  the disk normalization goes to zero. See Section \ref{sec:nthcomp}
  for the details of Model 2.\vspace{3mm}}
\label{fig:nthcompfit}
\end{figure*}

\subsubsection{Model 2: {\tt nthcomp}}
\label{sec:nthcomp}
In the interest of quantifying the robustness of our results to the
choice of hard X-ray component, we also explore the evolution of the
disk with several prescriptions for Comptonization, including 
{\tt compTT} and {\tt nthcomp} \citep*{Zdziarski96,Zycki99}. Here we
discuss the results from our modeling with {\tt nthcomp}. To use this
model, we set the seed photon distribution to be 
a disk blackbody and tie the seed temperature to the disk
temperature. This allows the disk temperature to be well-constrained
even when the disk normalization goes to zero (see below). The
resulting variations are qualitatively very 
similar to those measured with {\tt simpl}, and are shown in Figure  
\ref{fig:nthcompfit}. The overall goodness of fit is
$\chi^{2}_{\nu}=1.09.$ The power law index $\Gamma$ and the electron
temperature $kT_{\rm e}$ are strikingly similar to the corresponding
parameters in Model 1, and although the normalization of {\tt nthcomp}
is defined differently than $f_{\rm SC},$ these two parameters also
behave alike. See Table \ref{table:cont} for a
comparison of the fit parameters at $\phi=0.06$. 

Still, there are several noticeable distinctions between Model 1
and Model 2, which are primarily related to the
fact that with {\tt simpl}, the photons in the hard X-ray component
come directly from the disk component, while {\tt nthcomp} treats them
as distinct. For example, we find in Model 2 that our disk radii and
temperatures are somewhat lower, peaking at approximately 100 km and
1.8 keV instead of 110 km and 2.2 keV. Furthermore, during the hard
pulse, the spectrum is so completely dominated by the {\tt nthcomp}
component that the disk normalization (and radius) go to zero. This
phenomenon is certainly unphysical, so we ultimately prefer {\tt
simpl} (and its use of $f_{\rm SC}$ to track the pre-Comptonized disk)
to {\tt nthcomp} as a description of the spectrum. 

Nevertheless, the strong changes in the accretion disk and
Comptonization parameters around the hard pulse appear to lead to the 
same conclusion as Model 1. In both models, the hard pulse is
characterized by strong scattering, a relatively steep photon index
($\Gamma=2.5$ in Model 1, 3.2 in Model 2), and a population of
scattering electrons with significantly-reduced energy ($E_{\rm
  fold}=6$ keV in Model 1, $kT_{\rm e}=4-5$ keV in Model 2). These
numbers are not characteristic of the canonical X-ray hard
state. Therefore, while it is clear that Comptonization models do
provide a good description of the X-ray spectrum, our findings here
confirm our conclusion from Section \ref{sec:powspec} that it is
difficult to interpret the hard pulse as a brief X-ray hard state.

\begin{deluxetable}{lccc}
\tabletypesize{\scriptsize}
\tablecaption{X-ray Continuum Properties at $\phi=0.06$ (Hard Pulse)
\label{table:cont}}
\tablewidth{0pt}
\tablehead{
\colhead{Parameter}  & 
\colhead{Model 1}  &
\colhead{Model 2}  &
\colhead{Model 3}}
\startdata
\vspace{1mm}$R_{\rm in}$ (km) & $41\pm2$& $<12.4$&$22.1\pm0.4$\\
\vspace{1mm}$T_{\rm obs}$ (keV)& $2.12\pm0.08$& $1.74\pm0.02$&$2.19\pm0.03$\\
\vspace{1mm}$\Gamma$ & $2.86_{-0.07}^{+0.11}$& $3.29_{-0.29}^{+0.03}$&$>3.2$\\
\vspace{1mm}$f_{\rm SC}$ & $>0.99$& \nodata &$0.007_{-0.004}^{+0.027}$\\
\vspace{1mm}$N_{\rm nthcomp}$ & \nodata &$6.23_{-0.76}^{+0.03}$&\nodata \\
\vspace{1mm}$K_{\rm bremss}$  & \nodata& \nodata&$9.56\pm0.06$\\
\vspace{1mm}$E_{\rm fold}$ (keV)& $8.4_{-0.9}^{+1.4}$&\nodata &\nodata\\
\vspace{1mm}$kT_{\rm e}$ & \nodata & $4.70_{-0.07}^{+0.14}$&\nodata\\
\vspace{1mm}$kT_{\rm bremss}$ (keV)& \nodata& \nodata&$6.38_{-0.26}^{+0.05}$\\
\vspace{1mm}$F_{\rm 6.4~keV}$ & $0.027_{-0.009}^{+0.008}$& $0.012_{-0.007}^{+0.006}$&$0.051\pm0.008$\\
\vspace{1mm}$\sigma_{\rm 6.4~keV}$ (keV)& $<1.0$& $<1.2$&$0.8\pm0.2$\\
\vspace{1mm}$\chi^{2}/\nu$ & 21.9/43 &24.5/43&~~32.0/40
\enddata
\tablecomments{The hard pulse is the point of the $\rho$ cycle that
  has the most unusual spectral and timing properties, and coincides
  with the most dramatic changes in the continuum parameters. It
  therefore provides an ideal contrast between our models. The values
  reported here are averages of the two PCU combinations; the
  individual errors (90\% confidence ranges for a single parameter)
  have been added in quadrature with the standard deviation of the
  values for the two PCU sets. $\chi^{2}$ is calculated for 56 data
  points.}
\end{deluxetable}

\subsubsection{Model 3: Bremsstrahlung}
\label{sec:bremss}
We are now left with a small dilemma. The Comptonized disk models
provide a very good statistical description of the phase-resolved
X-ray spectra of the heartbeat state. However, several points indicate
the possible presence of a second hard component in the X-ray
spectrum. The most powerful argument in favor of such a component
comes from the X-ray spectra themselves. For more than 70\% of the
cycle, both the photon index and the electron temperature are
completely constant, and the scattering fraction varies smoothly. But
these parameters change dramatically during the hard pulse and then
return to ``normal.'' This suggests that there may be two
geometrically-distinct sources of hard X-rays (one roughly constant
and one variable). Additional evidence for a second component
comes from the power density spectra (Section \ref{sec:powspec}):
during the hard X-ray tail and the beginning of the slow rise, the
$rms$ level, the shape of the PDS, and the evolution of the QPO
frequency are all consistent with the properties of the
hard-state-like dips in GRS 1915+105. These dips are known to be
dominated by Compton scattering. The hard pulse
(Fig.\ \ref{fig:philc}b), on the other hand, exhibits a dip in $rms$ 
and there is no low-frequency QPO, which is typically associated with 
Compton-dominated intervals at high luminosity. This further indicates
that the mechanism producing the hard X-rays during the hard pulse may
be distinct from that in the hard X-ray tail and the slow rise. 
Finally, our timing analysis of 242 \textit{RXTE} observations of the
$\rho$ state (Neilsen et al.\ 2011, in preparation) and the
theoretical models of \citet{Nayakshin00} and \citet{JC05} support the
presence of a second component during the hard pulse that may be
related to plasma ejections from the inner disk. 

To test this idea, we replace the high-energy cutoff in Model 1 with a
thermal bremsstrahlung component to represent the emission from this
ejected plasma. The overall quality of the fit is comparable to the
other models ($\chi^{2}_{\nu}=1.12$), but because {\tt simpl} alone
provides a decent fit outside the X-ray pulses, the hard component is
occasionally over-determined. Nevertheless, this model presents a
physically-interesting alternative to the Compton-dominated models, so
we discuss it here briefly.

The accretion disk temperature and inner radius behave similarly to
the parameters shown in Figure \ref{fig:model1}, although we find
again that the disk normalization is smaller due to the
\begin{figure}[t]
\centerline{\includegraphics[width=0.49\textwidth]{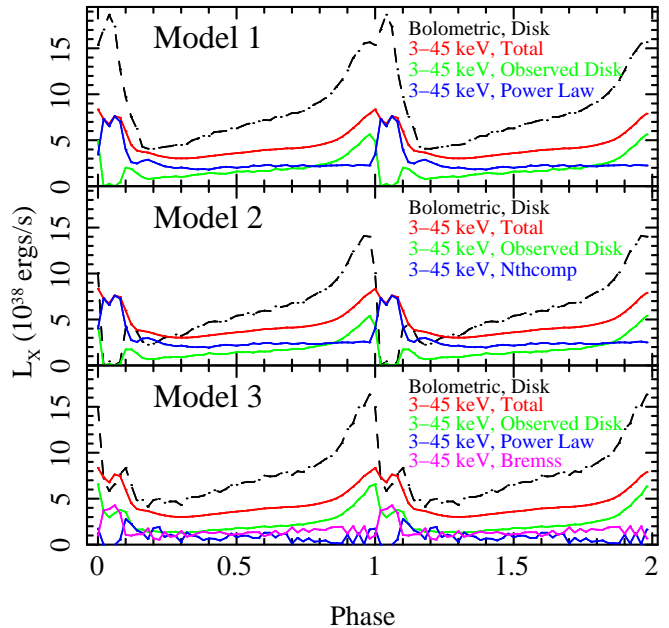}}
\caption{Luminosities of spectral components as functions of cycle
  phase. In the top panel, for Model 1, the bolometric disk luminosity
  (before scattering) is shown as a dot-dashed black line; we plot the
  3--45 keV total, observed disk, and scattered (power-law)
  luminosities as red, green, and blue solid lines, respectively. The
  middle panel is the same but for Model 2, with the luminosity of the
  {\tt nthcomp} component in blue. The bottom panel (Model 3)
  additionally shows the bremsstrahlung component in purple.}
\label{fig:lum}
\end{figure}
presence of
the bremsstrahlung component. But in light of the sudden changes in
the hard X-ray components during the hard pulse in Models 1 and 2, we
are primarily interested in the phase dependence of the bremsstrahlung
normalization and temperature in Model 3. Here, we find a sharp spike
in the bremsstrahlung normalization from $K=1-2$ at $\phi=0$ to a
value of $K=9.56\pm0.06$ at $\phi=0.06.$ 
%
%
The normalization
is given by \begin{equation}\label{eq:bremss}
K=\frac{3.02\times10^{-15}}{4\pi D^{2}}\int 
n_{e}n_{i}dV,\end{equation} where $D$ is the distance to GRS 1915+015,
$n_{e}$ and $n_{i}$ are the electron and ion number densities, and $V$
is the emitting volume. The bremsstrahlung temperature during this
spike is $kT_{\rm bremss}\sim5-7$ keV, which is consistent with the
evolution of the electron temperature in Models 1 and 2. Although the
exact physical interpretation of the hard X-ray component varies from
model to model, all three models require the sudden appearance of a
new population of electrons during the hard pulse, which coincides
with dramatic changes in the accretion disk. These electrons are
cooler than typical coronal electrons, but warmer than the disk. 


For comparison, we show the luminosities of the X-ray continuum
components in Figure \ref{fig:lum}, from which several points are
immediately clear. First, the hard X-ray luminosity is relatively
steady in all three models, with the exception of the hard pulse,
which is scattering dominated in Models 1 and 2 but 56\%
bremsstrahlung in Model 3. Second, the hard X-ray tail
($\phi=0.15-0.3$) is also dominated by scattering (here the disk
component constitutes $\lesssim30\%$ of the light in all
models). Finally, we can see that in all models, just before $\phi=0$
the bolometric disk luminosity (see also \ref{sec:mdot} for details)
reaches 80-90\% of the Eddington luminosity for GRS 1915+105 (a 14
$M_{\sun}$ black hole; $L_{\rm Edd}=1.8\times10^{39}$ ergs
s$^{-1}$). Model 1 actually tops out at $1.03L_{\rm Edd}.$ 

\subsubsection{Disk Variations and the Accretion Rate}
\label{sec:mdot}
The most obvious feature of our fits to the heartbeat state X-ray
continuum is the strong variability in the temperature and inner
radius of the accretion disk. Before turning to a theoretical
interpretation of the $\rho$ cycle (Section 5), we assess
the implications of our continuum fits for the mass accretion rate onto
the black hole. 
\begin{figure}
\centerline{\includegraphics[width=0.47\textwidth]{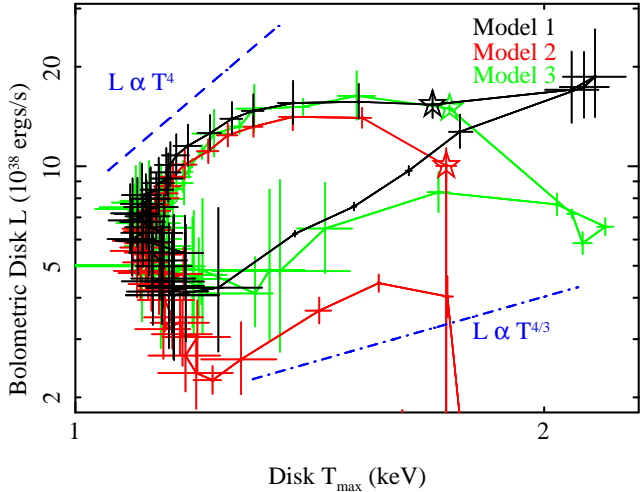}}
\caption{Accretion disk bolometric luminosity versus temperature for
  our continuum models (Model 1 in black, Model 2 in red, Model 3 in
  green). We use a star to mark the location of $\phi=0$ for both
  curves; motion around the loop is clockwise. We also overplot dashed
  and dot-dashed blue lines of $L\propto T^{4}$ and $T^{4/3},$  which
  are expected if the disk radius or the accretion rate is constant,
  respectively. Note that we have cut off Model 2 ({\tt nthcomp})
  where the disk luminosity goes to zero.}
\label{fig:lt4}
\end{figure}

Given the color correction factor $f$ and the measured inner radius
and temperature of the disk, we can calculate the disk mass accretion 
rate using equation (4) of \citet{Zimmerman05}:
 \begin{equation}\label{eq:tmdot}
T_{\rm obs}=\frac{f}{2.05}\left(\frac{3 G M \dot M}{8\pi R_{\rm in}^{3}
  \sigma} \right)^{1/4}, 
\end{equation} where $G$ is the gravitational constant, $M$ is the
black hole mass, and $\sigma$ is the Stefan-Boltzmann constant. We
evaluate the accretion rate as a comparative scale, noting that the
absolute value is quite uncertain, and that we assume no changes in
spectral hardening or the radiative efficiency of the disk.
All three models give very similar mass accretion rates and
bolometric luminosities. We find that $\dot M$ peaks just before the
spike in the disk temperature at $\phi=0,$ at a value $\dot
M\lesssim1.5\times10^{19}$ g s$^{-1}.$ After the peak, the mass
accretion rate drops sharply to $\dot M\sim(1-3)\times10^{18}$ g
s$^{-1}.$ For all models the evolution of $\dot M$ is similar to that
of the bolometric disk luminosities (Figure \ref{fig:lum}).

The evolution of the accretion disk component can also be seen in
Figure \ref{fig:lt4}, where we plot its bolometric luminosity versus
its temperature. During the $\rho$ cycle, GRS 1915+105 traces a
clockwise loop in the $L-T$ plot. In Models 2 and 3, the loop is
distorted because of the competition between the disk, {\tt nthcomp},
and bremsstrahlung components, but the behavior is otherwise
similar. Although we calculate each point on the plot ala
\citet{Zimmerman05}: 
\begin{equation}\label{eq:lt4}
L=73.9\sigma\left(\frac{T_{\rm obs}}{f}\right)^{4}R_{\rm
  in}^{2}=73.9~\sigma T_{\rm obs}^{4}\frac{N_{\rm disk}D^{2}}{\cos~i},
\end{equation} there is essentially no interval longer than a few
seconds over which $L\propto T^{4}.$ After $\phi=0,$ which is marked
with a star, the disk moves rapidly from its moderate temperature, high
luminosity state through a high temperature, moderate luminosity state
($\phi\sim0.06$) to a low temperature, low luminosity state
($\phi\sim0.2$). After the minimum, the luminosity rises at roughly
constant temperature (in our nomenclature, this interval is the slow
rise; see Fig.\ \ref{fig:philc}b). For reference, we overplot lines
of $L\propto T^{4}$, which results from a constant inner disk
radius, and $L\propto T^{4/3},$ which is expected for a constant
accretion rate. These lines broadly match the evolution of $L$ going
into and out of the slow rise. We will return to this plot, its
interpretation, and the question of the mass accretion rate in Section
\ref{sec:discuss}.
\begin{figure}[t]
\centerline{\includegraphics[width=0.49\textwidth]{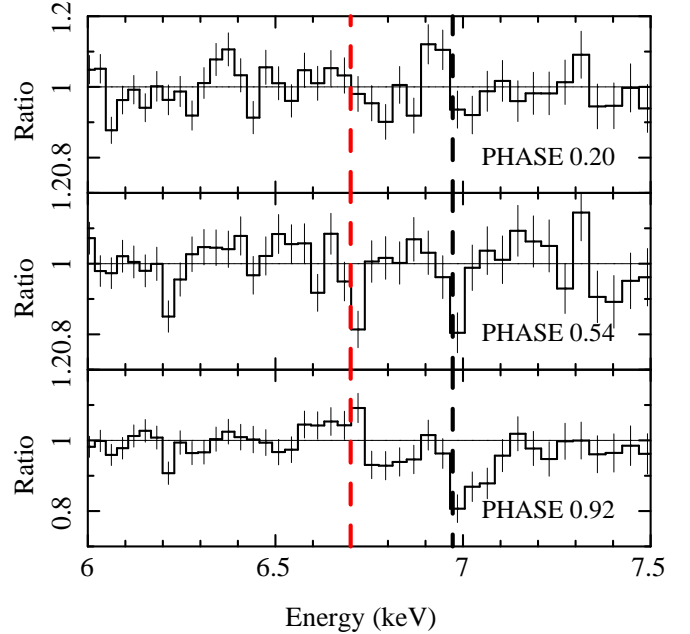}}
\caption{HETGS model/data spectra for three different phases of the
  heartbeat cycle with the rest energies of Fe\,{\sc xxv} (He$\alpha$:
  1s$^{2}$ -- 1s2p, 6.7 keV) and Fe\,{\sc xxvi} (Ly$\alpha$: 1s -- 2p,
  6.97 keV) marked in red and black, respectively. At $\phi=0.2$, near
  the cycle minimum, the iron absorption lines are weak. At
  $\phi=0.54,$ we detect noticeable features from Fe\,{\sc xxv} and
  Fe\,{\sc xxvi}, both blueshifted by $\sim500$ km s$^{-1}$. The
  feature at 6.2 keV may be Mn\,{\sc xxiv} He$\alpha$ or some other
  Doppler-shifted absorber. At  $\phi=0.92,$ the Fe\,{\sc xxv} line is
  gone, but there is some evidence for faster ($\sim5000$ km s$^{-1}$)
  and broader iron absorption in the Fe\,{\sc xxvi} line profile.}
\label{fig:HETGS_spec}
\end{figure}
\subsection{\textit{Chandra} HETGS}
\label{sec:hetgs}

While the \textit{RXTE} PCA provides high S/N broadband X-ray spectra
for measurements of the continuum, the \textit{Chandra} HETGS provides
an excellent high-spectral-resolution characterization of the observed
narrow lines in the soft X-ray band that is ideal for studying the
accretion disk wind and its dynamical evolution.

One of the most significant challenges for fast phase-resolved grating 
spectroscopy is choosing time and phase intervals to maximize the
detectability of interesting features and their possible
variations. For ease of comparison with the PCA spectra, we restrict
our attention here to the $\sim20$ ks of the \textit{Chandra}
observation that bracket the \textit{RXTE} data. This choice maximizes
the S/N in the Fe\,{\sc xxv} line, which increases over the course of
our observation, relative to Fe\,{\sc xxvi} (see Section  
\ref{sec:wind}). For our phase-dependent spectral analysis, we extract
50 HEG spectra evenly spaced in phase. In order to achieve sufficient
S/N, we set the phase width of each spectrum to $\Delta\phi=0.2$. This
is essentially a sliding box window, so there is some overlap between
consecutive spectra, but as shown by \citet{Schulz02}, this method is
quite suitable for line variability studies. As discussed in Section
\ref{sec:chandraobs}, because of calibration uncertainties we model
the \textit{Chandra} X-ray continuum with a polynomial fit (instead of
the physical models of Section \ref{sec:pca}). Polynomials accurately
characterize the local continuum, so they effectively isolate narrow
absorption lines. See Figure \ref{fig:HETGS_spec} for three example
residual spectra.
\begin{figure}
\centerline{\includegraphics[angle=270,width=0.49\textwidth]{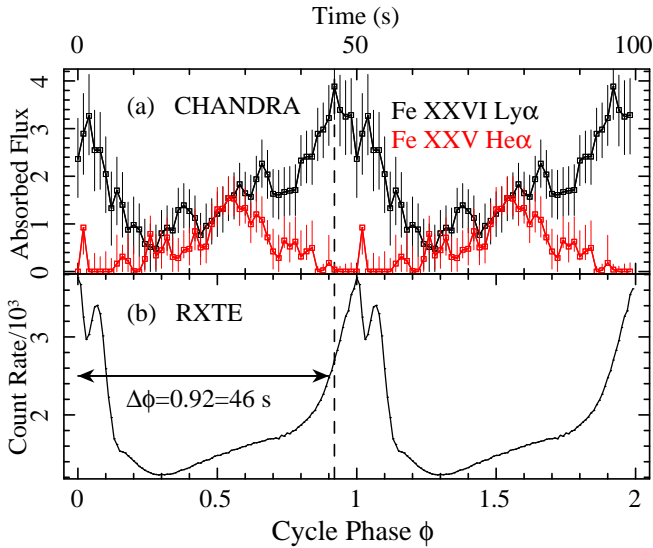}}
\caption{(a) Measured flux (in units of $10^{-3}$ photons s$^{-1}$
  cm$^{-2}$) in the Fe\,{\sc xxv} (red) and Fe\,{\sc xxvi} (black)
  absorption lines as a function of $\rho$ cycle phase. The H-like
  iron line clearly tracks the X-ray continuum, although the peak line
  flux follows the peak X-ray luminosity with a delay
  $\Delta\phi=0.92\approx46$ seconds. In contrast,
  the He-like line only appears briefly, growing in strength with
  Fe\,{\sc xxvi} until $\phi\sim0.54,$ where it starts to 
  fade. 
  (b) The full band \textit{RXTE} PCA phase-folded lightcurve, for
  ease of tracking changes in the absorption lines observed with
  \textit{Chandra} (top panel).}  
\label{fig:philine}
\end{figure}

A complete search for variable features at all velocities and
ionization levels with any arbitrary phase dependence is beyond the
statistical scope of this paper. Since we are mainly interested in the
known accretion disk wind in GRS 1915+105 (\citealt{L02};
\citetalias{NL09}), we focus on the observed narrow iron absorption
lines with velocities ranging from $\sim500$ to 2000 km s$^{-1}$ (see,
e.g.\ the middle panel of Figure \ref{fig:HETGS_spec}). We detect
both Fe\,{\sc xxvi} (Ly$\alpha$: 1s -- 2p, 6.97 keV) and Fe\,{\sc xxv}
(He$\alpha$: 1s$^{2}$ -- 1s2p, 6.7 keV). We use Gaussian fits to measure
the parameters of these iron absorption lines at each phase (see Table
\ref{table:abs}). To decouple our results from possible variations in
the wind dynamics, which could manifest as changing line widths and
thus affect the apparent line flux, we fix the intrinsic line width at
200 km s$^{-1}.$ This is a typical orbital speed in the outer disk,
and matches the turbulent line width found by \citet{U09} in their
study of the $\phi$ state. When we detect both Fe\,{\sc xxv} and
Fe\,{\sc xxvi}, we can fit for a 90\% confidence upper limit on the
line width of 800 km s$^{-1}.$ 

Finally, we note the possible presence of additional lines that vary
with phase, including a line at 6.2 keV that may be Mn\,{\sc xxiv}
He$\alpha$ and an unidentified feature at 4.29 keV. Without secure
identifications it is unclear if they are part of the iron absorber
or if they represent a different dynamical component, such as an
acceleration zone of the wind or a signature of infall. Both have
maximum fluxes around $2.0\pm0.5$ photons s$^{-1}$ cm$^{-2};$ the
corresponding equivalent widths are 2.5 eV and 6.4 eV,
respectively. Since we cannot clearly identify their origin we do not
consider them further here, beyond noting that their phase dependence
is somewhat similar to that of the iron absorber.


\begin{deluxetable*}{clccccc}
\tabletypesize{\scriptsize}
\tablecaption{X-ray Absorption Lines in the $\rho$ State of GRS 1915+105
\label{table:abs}}
\tablewidth{0pt}
\tablehead{
\colhead{}  & 
\colhead{}  &
\colhead{$E_{0}$}  &
\colhead{$E_{\rm obs}$}  &
\colhead{$\Delta v_{\rm shift}$}  & 
\colhead{}  & 
\colhead{$W_{0}$} \\
\colhead{Phase} & 
\colhead{Line}  &
\colhead{(keV)}  &
\colhead{(keV)}  &
\colhead{(km s$^{-1}$)}  &
\colhead{Flux}  &
\colhead{(eV)}}
\startdata
\vspace{1mm}0.20&Fe\,{\sc xxvi} Ly$\alpha$&6.966 &$7.01_{-0.01}^{+0.02}$&$-1700_{-1000}^{+500}$&$<1.6$&$>-8.6$\\
\vspace{1mm}&Fe\,{\sc xxv} He$\alpha$&6.700 &6.711\tablenotemark{a}&$-490$\tablenotemark{a}&$<0.5$&$-0.1_{-2.6}^{+0.1}$\\
\vspace{1mm}0.54&Fe\,{\sc xxvi} Ly$\alpha$&6.966 &$6.976_{-0.009}^{+0.018}$&$-430_{-750}^{+390}$&$1.5\pm0.5$&$-11.2_{-3.4}^{+3.5}$\\
\vspace{1mm}&\textbf{Fe\,\textsc{xxv} He}{\boldmath $\alpha$}&6.700 &$6.711_{-0.008}^{+0.010}$&$-490_{-420}^{+370}$&$1.5_{-0.3}^{+0.4}$&$-10.1_{-3.0}^{+2.2}$\\
\vspace{1mm}0.92&\textbf{Fe\,\textsc{xxvi} Ly}{\boldmath $\alpha$}&6.966 &$7.004_{-0.008}^{+0.001}$&$-1610_{-50}^{+360}$&$3.9\pm0.7$&$-14.1_{-2.4}^{+2.5}$\\
\vspace{1mm}&Fe\,{\sc xxv} He$\alpha$&6.700 &$6.711\tablenotemark{a}$&$-490\tablenotemark{a}$&$<0.1$&~~~$>-0.5$
\enddata
\tablecomments{Errors quoted are 68\% confidence ranges for a single
  parameter. Phase: line parameters are reported for three different
  phases of the cycle; $E_{0}:$ rest energy; $E_{\rm obs}:$ measured
  energy; $\Delta v_{\rm shift}:$ measured Doppler velocity; Flux:
  measured absorbed line flux in units of 10$^{-3}$ photons s$^{-1}$
  cm$^{-2};~W_{0}:$ line equivalent width; $\phi_{\rm max}:$ phase of
  maximum line flux. Boldface indicates the phase of maximum flux for
  that ion.}
\tablenotetext{a}{Since the Fe\,{\sc xxv} line is not detected at this
  phase, we fix its energy at the value measured at $\phi=0.54$ to
  place limits on the line flux and equivalent width.\vspace{5mm}}
\end{deluxetable*}

It is clear from both Table \ref{table:abs} and Figure
\ref{fig:HETGS_spec} that the ionization parameter $\xi$ of the wind
changes substantially during the cycle. Table \ref{table:abs} shows
that Fe\,{\sc xxv} and Fe\,{\sc xxvi} peak with similar equivalent
widths at very different phases ($\phi=0.54$ vs.\ $\phi=0.92$), while
a comparison of the middle and bottom panels of Figure
\ref{fig:HETGS_spec} shows that the Fe\,{\sc xxvi} absorption line is
strong even when the Fe\,{\sc xxv} line has disappeared. In the rest
of this section, we focus on these two later phases as representative
of the variations in disk wind absorption in this state. 

This ionization evolution is even more obvious in Figure
\ref{fig:philine}. The Fe\,{\sc xxv} absorption line appears to track
the Fe\,{\sc xxvi} line in the interval $\phi=0.2-0.54,$ but it fades
rapidly outside this interval and is not significantly detected for
the rest of the cycle. The Fe\,{\sc xxvi} absorption line, however,
continues to grow until $\phi=0.92.$ The peak flux in this line is
more than twice that measured from the average spectrum; Monte Carlo
simulations indicate that this increase is significant at 96\% confidence.

It is also evident in Figure \ref{fig:philine} that the Fe\,{\sc xxvi}
absorption line evolves similarly to the PCA count rate.  
Under the assumption that the line variability is a response
to the X-ray continuum impinging on the wind, it appears that the peak
line flux follows the X-ray flux with a delay $\Delta\phi\approx
0.92=46$ seconds. This is actually rather long: the semimajor axis of
GRS 1915+105 is $a\sim250$ lt-s \citep*{G01}, and previous estimates
have placed the wind near $r\sim10$ lt-s from the black hole
(\citealt{L02,U09}; \citetalias{NL09}). It may then be reasonable to suspect
that $\Delta\phi$ is mainly due to continued increases in the
ionization of the absorbing gas. We will show in Section
\ref{sec:diskwind} that changes in the gas density are also
important.

Additionally, we note that our measurements of the absorption line
centroids provide marginal evidence for variations
in the blueshift of the wind during the cycle. At $\phi=0.54,$ we
detect Fe\,{\sc xxvi} at $v\sim-430^{+390}_{-750}$ km s$^{-1};$ by
$\phi=0.92,$ the velocity has increased to $v\sim-1600_{-50}^{+360}$
km s$^{-1}.$ This apparent acceleration is significant at 96\%
confidence, and it occurs in combination with the appearance of 
a blue wing to the Fe\,{\sc xxvi} profile (see
Fig.\ \ref{fig:HETGS_spec}). We then take the velocity and line width 
as some evidence of variability in the dynamics of the disk wind. In
Section \ref{sec:diskwind}, we discuss the photoionization evolution,
dynamics, and the 46-second delay in the context of wind formation
scenarios and the disk-wind-jet connection.
\section{DISCUSSION}
\label{sec:discuss}
In this section, we explore the significance of our results
from Sections \ref{sec:timing} and \ref{sec:spectra}. Specifically, we
will address our findings of (1) strong, fast variations in the inner
radius of the accretion disk, (2) bremsstrahlung emission during the
hard pulse, and (3) clear changes in the absorption lines from the
accretion disk wind on timescales of 5 seconds. 

While strong variations in the X-ray spectrum of GRS 1915+105 are
routinely observed, the magnitude and short timescale of the spectral
variability in the heartbeat state bears repeating, especially given
its cyclic nature. In the rising phase of the cycle, the bolometric
disk luminosity changes from about ($5$ to $16)\times10^{38}$ ergs
s$^{-1}$ in roughly 35 seconds; during the decay phase, the X-ray
luminosity decreases by as much as $\dot  L_{\rm X}
\lesssim5\times10^{38}$ ergs s$^{-2}.$ In other words, we find that
the black hole halts a burst reaching $\sim80-90\%$ of its Eddington
luminosity in just a few seconds. As discussed in Section 4, the
luminosity variations are associated with substantial changes in the
inner radius and temperature of the accretion disk.

The physical changes in this bizarre oscillation are represented in
the phase variation of the X-ray continuum, fluorescent emission
lines, and the hot iron absorption lines of the accretion disk
wind. In the following subsections, we discuss our results on
phase-resolved spectra and power spectra to understand the origin,
dynamics, and accretion geometry of the $\rho$ state and to explore in
detail the apparent coupling between the accretion disk wind and the
X-ray oscillation. \textit{We describe our model for the physics of
  the heartbeat state, starting with variations in the accretion rate 
  in the inner disk, and tracing their influence on the accretion 
  dynamics all the way to the outer edge of the disk.} 

\subsection{X-ray Continuum}
\label{sec:continuum}
\subsubsection{$\dot M$ Variations and Z Source-like Behavior}
\label{sec:zsource}
The most obvious properties of the heartbeat state (based on
our X-ray continuum fits) are the observed strong variability in the
luminosity, temperature, and inner radius of the accretion disk
(Figs.\ \ref{fig:model1}--\ref{fig:lt4}). Specifically, our analysis
shows that during the slow rise, the disk radius gradually grows, then
drops sharply and rebounds during the X-ray peaks. Meanwhile, the disk
temperature appears constant and then spikes rapidly.

\citeauthor*{Lin09}~(2009, hereafter LRH09) discovered similar~ variations
in the radius, temperature, and luminosity of the accretion disk in the
accreting neutron star XTE J1701-462 as it evolved from a Z-source to
an atoll source. Z sources are accreting neutron stars that display
three branches in the shape of a ``Z'' in their color-color
diagrams. LRH09 found that in the ensemble of observations in the
vertex between the flaring branch and normal branch, the inner radius
of the disk increased with luminosity above $L\sim0.2L_{\rm Edd},$
while the disk temperature remained roughly constant. They interpreted
this result as an indication of a local Eddington limit
\citep{Fukue04,Heinzeller07} in the inner disk. Local Eddington effects arise
because the radiation forces and gravitational forces depend
differently on radius in thin disks, so that there exists a critical
radius where radiation pressure can truncate the disk. Inside this
critical radius, a significant fraction of the disk is expelled by
radiation pressure. The resulting spectrum is the same as that of a
standard thin disk for energies $E\gtrsim k_{B}T_{\rm crit}.$ It is
possible that the same process is at work during the slow rise of the
$\rho$ cycle in GRS 1915+105, i.e.\ when the radius increases with
luminosity at roughly constant $T$ (Fig.\ \ref{fig:lt4}).

LRH09 also showed that on the XTE J1701-462 flaring branch, the 
luminosity is proportional to $T^{4/3},$ which is expected if the
inner radius of the disk changes while the accretion rate remains
constant. They concluded that the flaring branch is an instability in
which the disk temporarily slips inwards but is quickly driven back to
the equilibrium position seen at the vertex position on that same
day. In their interpretation, this equilibrium position is set by the
local Eddington limit. For comparison, we plot a line of $L\propto
T^{4/3}$ in Figure \ref{fig:lt4}. It can be seen that in GRS 1915+105,
the disk very roughly approximates this scaling as it both approaches
and exits the slow rise of the $\rho$ cycle. We also show $L\propto
T^{4}$; the two lines bracket the behavior of the accretion disk
during the $\rho$ cycle. These rough scalings seem to indicate
that some changes in the disk may occur at $\sim$constant accretion
rate, especially after $\phi=0$. 

Thus, mapping LRH09's prescription for neutron stars onto GRS
1915+105, we can understand the gradual increase in the disk
radius during the slow rise of the heartbeat state as a local
Eddington phenomenon (our measured radii and temperatures are
consistent with those predicted by \citealt{Fukue04}). The disk slowly
expands with an increasing accretion rate. After the accretion rate
reaches a maximum the disk first falls in at $\sim$ constant accretion
rate and then cools rapidly while the accretion rate drops, setting the stage
for the next cycle. It should be noted that the analogy between the
XTE J1701-462 lower vertex/flaring branch and the $\rho$ cycle is not exact, since the variations in GRS 1915+105 are much faster, and XTE
J1701-462 does not trace out a loop in the $L-T$ diagram. Furthermore,
only GRS 1915+105 shows the rapid catastrophic changes seen in the
disk and the accretion rate around $\phi=0$.

Nevertheless, it is very exciting to see such similar accretion
disk behavior between a neutron star system and a black hole system;
the commonalities suggest that the $\rho$ cycle may share or mimic some
properties of a Z source. This particular comparison seems reasonable, 
since Z sources are known for high accretion rates (see LRH09 and
references therein), and GRS 1915+105 may have the highest accretion
rate of any known black hole transient. These similarities also
support the conclusion of \citet*{Done04} that GRS 1915+105 may be unique
merely for its high accretion rate.

\subsubsection{On the Role of Radiation Pressure}
\label{sec:rad}
Our interpretation that the local Eddington limit in the inner
accretion disk may produce the slow rise during the $\rho$ cycle (see
Section \ref{sec:zsource}) poses an intriguing question as to the role
of radiation pressure in the heartbeat state. We now have claims of
two very different radiation pressure phenomena in the $\rho$ state: a
local Eddington limit and (historically) the Lightman-Eardley 
radiation pressure instability (RPI). The RPI is commonly invoked,
with some success, to explain the strong limit cycle oscillations in
GRS 1915+105 (\citetalias{TCS97}; \citealt{Nayakshin00}, hereafter
NRM00; \citealt{JC05}, 
hereafter JC05). These processes are physically independent but not
necessarily conflicting. Here we consider the question: what is the
true role of radiation pressure in the $\rho$ state? Does it produce
limit-cycle oscillations by making the disk (intact at constant
$R_{\rm in}$) thermally/viscously unstable, or does it actually push
out the inner radius of the disk over timescales of $\sim25$
s $\approx4000$ dynamical times? Or are some elements of both models
require to explain the observations? To answer this question, we need
to understand how the observed changes in the radius, temperature, and
luminosity of the accretion disk might be related to these physical
mechanisms.

For the local Eddington limit, $R_{\rm in}$ increases with $L$ at
constant $T$. For half of each observed $\rho$ cycle (the slow rise;
Fig.\ \ref{fig:philc}b), the disk luminosity and radius appear to grow
at $\sim$constant $T$. Alone, however, a local Eddington limit cannot
produce an oscillation or a loop in the $L-T$ diagram. Some other
mechanism is required to cyclically vary the accretion rate for a
complete description of the heartbeat state. 

The question is more subtle as it relates to the RPI. Does the RPI act
in concert with a local Eddington limit to produce the $\rho$ state,
or can it mimic an interval of increasing $R_{\rm in}$ at constant
$T$? Since time-dependent disk simulations of the RPI (e.g.\ NRM00 and
JC05) have not produced synthetic spectra, the question is also
difficult to answer. For preliminary considerations, we examine the
dynamics of the accretion disk in their simulations. 

Physically, the RPI occurs when the inner disk becomes unstable due to
radiation pressure. This drives a limit cycle involving an oscillation
of the local accretion rate in the inner disk. This can be interpreted
as a `density wave', which originates around $R_{\rm dw}=20-30  
R_{\rm g}.$ Here $R_{\rm g}$ is the gravitational radius (21 km for
GRS 1915+105). As the density wave moves through the disk, it
partially evacuates a gap in the disk. Inside and outside the gap, the
disk surface density is enhanced (see Figure 4 in both NRM00 and
JC05). The inner density enhancement peaks around $R=5R_{\rm g}
\approx105$ km. The similarity of this estimate to our measured radii
is striking. During the decay of the cycle, this excess density moves
inwards quickly (and our measured radii decrease sharply). These
results provide a hint as to why the apparent disk radius might change
rapidly in the pulses. 

It is unclear, however, if the RPI can mimic our observation of
increasing $R_{\rm in}$ at constant $T$ during the slow rise. Our
local Eddington interpretation implies that the true edge 
of the disk moves outwards with $L$, while NRM00 and JC05 have
$R_{\rm in}$ fixed at $2~R_{\rm g}.$ On the other hand, the disk
surface density and temperature are increasing and decreasing
functions of radius, respectively, and both groups find a short hot
state and an extended ``cold'' state. For these reasons, it seems
potentially possible for their time-dependent thin disk models to
mimic an expanding-radius disk at constant $T$.

Thus the role of radiation pressure in the heartbeat state requires
some additional model analyses. A local Eddington limit in the disk
does a good job explaining the slow rise, but cannot explain the
pulses or the existence of an oscillation. The RPI is a viable
explanation for the oscillation, but it is unclear if it can either
act alone or jointly with the local Eddington limit to reproduce our
observations of the slow rise. Future time-dependent disk simulations
will be able to resolve this issue by calculating model spectra that
can be compared with the observations. For now, it 
seems that our observed disk radii and temperatures provide some
support for both the RPI and a local Eddington limit at work in GRS
1915+105.

\subsubsection{Ejections, Jets, and the Fate of the Inner Disk}
\label{sec:eject}
Another exciting question posed by the observed rapid oscillations in
the radius, temperature, and mass accretion rate in the disk concerns
the ultimate fate of the excess material in the inner disk. Does it
simply fall into the black hole, or is it ejected from the inner disk
and therefore potentially observable? We have not directly observed
any ejection events, but the mysterious hard pulse in the X-ray
lightcurve still lacks a satisfying interpretation. It is even
peculiar theoretically, since RPI simulations produce only
single-peaked cycles. Is this short burst of hard X-rays related to
the ejection of material from the inner disk, or does it have an
explanation in terms of standard black hole states?

The association of the hard pulse with a typical soft-to-hard state
transition appears to be ruled out, since the spectral and
timing properties of the hard pulse are not remotely consistent with
typical hard states for either GRS 1915+105 or black hole binaries in
general. That is, although we are able to successfully model its
X-ray spectrum with disk/Comptonization models, the resulting
parameters are inconsistent with the canonical hard state, and
imply the sudden appearance of a low-temperature, high-optical-depth
cloud of electrons ($\tau\sim5$ with {\tt nthcomp}). 

Before going ahead, we may ask if the appearance of these cooler
electrons can be explained simply by an increase in their cooling
rate. If synchrotron cooling is important, additional cooling could be
accomplished by an increase in the electron density or magnetic field
strength. If inverse Compton cooling dominates, then an increase in 
cooling could result from the observed increase in the X-ray flux or
from an increase in the electron density. Detailed radiative transfer
calculations would be necessary to determine the energetics precisely,
but it is unclear in the Compton cooling case why the cooling is is
abrupt, while the flux changes smoothly. The simplest explanation in
both cases is a sudden increase in the electron density. This
justifies our use of a bremsstrahlung component to model the hard
pulse, and provides us with an observational basis for discussing
plasma ejections from the inner disk. 

Additional support for plasma ejections comes from  the theoretical
work of NRM00 and JC05, who found that the X-ray states of GRS
1915+105 are reproduced with higher fidelity when ejection processes
are included.  NRM00 allowed a fraction of the total accretion power
to be channeled into a jet with Lorentz factor $\gamma\sim3.$  JC05,
modeling the $\rho$ state specifically, allowed the disk to evaporate
into the corona at the peak of the cycle.

It should be noted that these models (see also \citealt{J00}) produce
plasma ejections when the X-ray luminosity is greatest, and are
therefore at odds with the conclusions of \citet{K02}, who
argued that jet production is essentially a continuous process that
occurs during hard faint intervals (State C in the classification of
\citetalias{B00}). However, the $\rho$ state is only known to exhibit low-level
radio emission, and it is unclear if this is a `baby' jet with a short
duty cycle or some other stable but faint radio activity. Thus there
may be no observational conflict with \citet{K02} for the $\rho$ 
state. Future high time-resolution radio/X-ray observations
(e.g.\ with the EVLA and \textit{Chandra}) should probe the precise
nature of the radio emission in such states.

In any case, both groups predict plasma ejections near $\phi=0,$ and
our models implies the sudden appearance of some cold plasma around
the same phase. With the reasonable assumption that these two points
are related, we proceed to estimate how much material is ejected
from the disk. For this, we can use equation (\ref{eq:bremss}), which
relates the bremsstrahlung normalization to the volume and electron
density in the emitter. Because the bremsstrahlung 
pulse begins just after the drop in $R_{\rm in}$ at $\phi\sim0,$ we
assume that $R_{\rm  bremss}\sim\max(R_{\rm in})\sim 97$ km, so that
the emitting volume is $V_{\rm max}\sim4\times 10^{21}$ cm$^{3}.$ In
other words, we assume that some bremsstrahlung-emitting material is
ejected from the inner disk, filling a sphere of radius 97
km. Assuming the particle density is constant over the sphere and that
the material is fully ionized (with ISM abundances from
\citealt{Wilms00}), we find an implied ion density $n_{\rm i}\sim10^{20}$
cm$^{-3}$ and ion mass $m_{\rm i}\sim10^{18}$ g. For a 
Shakura-Sunyaev disk with $\dot M\sim10^{19}$ g s$^{-1},$ the disk
mass interior to 97 km is of order $10^{21}$ g, $\gtrsim1000$ times
larger than the mass of this bremsstrahlung cloud. Since JC05 argued
that as much as $10\%$ of the disk accretion rate may evaporate into
the corona, our requirement that 0.1\% of the disk mass contribute to
the bremsstrahlung pulse seems reasonable.

Also of interest is the possible production of a collimated jet at some
point in the $\rho$ cycle. In Section
\ref{sec:powspec}, we considered the extent to which the X-ray minimum
of the $\rho$ state, or even the slow rise, could be considered a hard
state, with the purpose of identifying those phase intervals where we
might expect to see some signatures of jet production. Our examination
of the phase-dependent PDS and X-ray continuum indicates that the end
of the hard X-ray tail, $\phi=0.26-0.36,$ has timing and spectral
properties that are consistent with an X-ray hard state. Thus if
there is any jet activity, we might expect it to have a duty cycle of
order 10\%. In this window around the count rate minimum, a hard state
(and jet) may temporarily coexist with the disk. This conclusion is
particularly interesting in light of the possibility of strong
synchrotron cooling during the hard pulse, but a radio detection of a
synchrotron spectrum would be required to verify and connect these two
phenomena. 

Previously, we reported the Ryle telescope radio flux near
this observation to be $S_{\rm 15GHz}\lesssim5$ mJy \citep{PF97,NL09},
which is comparable to the sensitivity of the Ryle telescope. Thus we
cannot draw any robust conclusions about a jet. We note, however, that
this is averaged over nearly a full cycle; if the actual radio flux is
close to this upper limit, then the ``true'' jet strength could be an
order of magnitude higher. This would be typical for a $\chi$ state
jet in GRS 1915+105 \citep{F99,K02}. Since a short-lived jet is
plausible, future sensitive radio observations of the $\rho$ state may
place strong constraints on jet evolution around black holes. 

To summarize, it appears that our Comptonized disk plus bremsstrahlung
model leads to a physically-sensible observational interpretation of
the $\rho$ state in the context of plasma ejections. We find that the
double-peaked X-ray lightcurve is the result of two processes: (1)
large, cyclic variations in the mass accretion rate in the inner disk,
leading to an increase in the disk temperature (the soft pulse), and
(2) the ejection/evaporation of a small fraction of the inner disk,
which produces a brief burst of bremsstrahlung as it propagates into
the corona (the hard pulse). The hard X-ray emission could also simply
be scattering in a Compton-thick cloud, but the appearance of this
cloud is still best explained by material evaporating from the
disk. This ``steam'' from the disk does not qualify as a jet, but it
is nevertheless an exciting signature of plasma ejections during the
heartbeat cycle.

\subsection{Iron Emission Line}
\label{sec:feline}
\begin{figure}
\centerline{\includegraphics[width=0.49\textwidth]{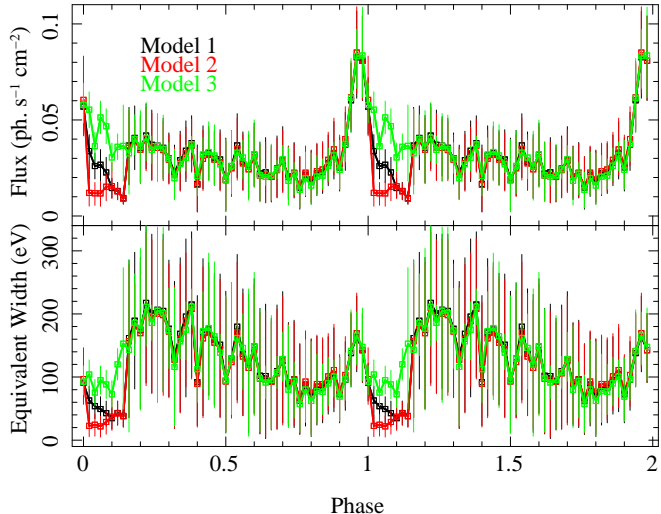}}
\caption{Flux (top) and equivalent width (bottom) of the broad iron
  emission line as a function of phase. The line flux is roughly
  constant during most of the cycle and spikes during the two pulses,
  while the equivalent width oscillates slowly.}
\label{fig:feline}
\end{figure}

Here we briefly consider the broad iron line seen in our \textit{RXTE} 
spectra (Figure \ref{fig:feline}). In our models, the line flux slowly
decreases for most of the cycle and the equivalent width
oscillates slowly, spiking during the soft pulse. Models 1 and 2 show
a dip in the line strength during the hard pulse, but the reality of
this feature is unclear. In the standard interpretation
(\citealt{Fabian89}; \citealt*{Matt93,RN03}, and references therein), the broad line is a
reflection feature from the inner accretion disk. As shown in Section
\ref{sec:pca}, the innermost regions of the disk fluctuate cyclically
during the $\rho$ state, so the fact that the line flux is
almost constant indicates that some geometrical effects from the
oscillation may be influencing the illumination of the disk. For
example, the increase of the disk scale height with luminosity may
shield outer regions of the inner disk from some of the X-ray
luminosity, allowing the reflected line flux to stay the same. 

On the other hand, if the line originates at the inner edge of the
disk, then the constant line flux might be caused by a combination of
increased illumination and decreased reflecting area. We note that the
line equivalent width looks rather like the {\tt simpl} scattering
fraction, so it seems likely that the changes in the hard X-ray
component are literally reflected in the iron line. But without
detailed theoretical models of the disk spectrum and the iron line
profile, a robust determination of its location is not possible.
Thus it would be difficult to use this broad line to place a constrain
on the black hole spin. In other states like the $\chi$ state, where
the X-ray variability and its effects on the disk structure are not so
important, estimates of the spin parameter $a_{*}$ may be both more
feasible and more meaningful \citep{Martocchia02,Blum09}.

\subsection{Accretion Disk Wind}
\label{sec:diskwind}
One of the main goals of this work, as a follow-up to our discovery of
a wind-jet interaction in Paper 1, is to understand the complex
relationship between the fast X-ray variability and the physics of the
accretion disk wind, i.e.\ its origin, dynamics, ionization,
structure, and short-timescale evolution. This includes determining
whether or not the structure of 
the wind is constant in time. One could easily imagine a scenario in
which the wind is unaffected by the limit cycles in the inner disk,
e.g.\ if it is launched from the outer disk by simple X-ray heating
from the phase-averaged X-ray luminosity or by MHD processes (see,
e.g.\ \citealt{M06a,M08}). Or perhaps there is a simple relationship
between the wind and the variability, where the wind forms
independently but, as it rises off the disk, is photoionized by  
the strong quasi-periodic X-ray variability.

In this section, we exploit the simultaneous strong variability in the
accretion disk wind (Figure \ref{fig:philine}) and the X-ray continuum
to argue that neither of these scenarios provides a satisfactory
description of the accretion disk wind physics in GRS 1915+105. We use
photoionization arguments to probe the accretion dynamics and plasma
conditions in the outer disk, and conclude that there must be a
powerful coupling between the formation of accretion disk wind and 
the X-ray luminosity. We argue that photoionization alone is
insufficient to produce the observed variability in the wind.

\subsubsection{Plasma Conditions and Ionization Balance}
In Section \ref{sec:hetgs} we showed that the absorbed fluxes in both 
Fe\,{\sc xxv} and Fe\,{\sc xxvi} vary with $\phi.$ Given the
variability in the X-ray absorber, we treat the disk wind as though it
is not (in general) in photoionization equilibrium. However, it is
reasonable to assume that the observed maxima in the absorbed fluxes
occur when the instantaneous photoionization and recombination rates
are momentarily equal. Here, under this assumption, we estimate the
gas density in the wind.

Neglecting collisional ionization, this temporary ionization balance
can be expressed \citep{Liedahl99}
\begin{equation}\label{eq:ionbal}
\frac{1}{4\pi R^2}~\Phi_{i}n_{i}=\alpha_{i+1}n_{\rm e}n_{i+1},
\end{equation} 
where $R$ is the distance to the continuum source, $n_{\rm e},~n_{i},$ and
$n_{i+1},$ are the electron density and number density of iron in
charge states $i$ and $i+1,~\alpha_{i+1}$ is the recombination rate
into charge state $i$ (out of charge state $i+1$), and $\Phi_{i}$ is
the ionizing spectrum integrated over the photoionization
cross-section $\sigma_{i}$ for charge state $i:$ 
\begin{equation}\label{eq:int}
\Phi_{i}=\int_{\chi_{i}}^{\infty}\varepsilon^{-1}\sigma_{i}(\varepsilon)
L_{\varepsilon}d\varepsilon.
\end{equation} 
Here $L_{\varepsilon}$ is the monochromatic luminosity and $\chi_{i}$
is the ionization threshold. Given a set of ion fractions
(or $n_{i}$ and $n_{i+1}$), an ionizing continuum, and
appropriate recombination rates, the ionization parameter \citep{Tarter69} 
\begin{equation}\label{eq:ion}
\xi=\frac{L_{\rm X}}{n_{\rm e}R^2}
\end{equation}
is determined by equations (\ref{eq:ionbal}) and (\ref{eq:int}). 

In general, the ion fractions depend on the ionizing continuum, the
ionization parameter, and the overall density \citep{Kallman01}. However,
if the absorption lines are unsaturated, we have an additional
constraint from the linear part of the curve of growth, which relates
the equivalent width of a spectral line to the ionic column density: 
\begin{equation}\label{eq:cog}
\frac{W_{\lambda}}{\lambda}=\frac{\pi e^{2}}{m_{e}c^{2}}N_{i}\lambda
f_{ji}.
\end{equation}
Here $W_{\lambda}$ is the line equivalent width in m\AA, $\lambda$ is
the wavelength, $m_{e}$ is the electron mass, $c$ is the speed of
light, $N_{i}$ is the column density of charge state $i,$ and $f_{ij}$
is the oscillator strength of the relevant transition. 

At $\phi=0.54,$ the Fe\,{\sc xxvi} line is almost certainly
unsaturated because it continues to increase in both absorbed flux
and equivalent width for another 20 seconds (for saturated lines, the
flux and equivalent width grow very slowly until the optical depth
is $\gtrsim1000$). For this reason, it seems likely that the 
Fe\,{\sc xxv} line is also unsaturated. Using our measured equivalent
widths and oscillator strengths from the {\sc xstar}
\citep{Kallman01,Bautista01} 
database, we evaluate equation \ref{eq:cog} for Fe\,{\sc xxv} and
Fe\,{\sc xxvi}. We find that $n_{\rm XXVI}/n_{\rm XXV}\approx 
N_{\rm XXVI}/N_{\rm XXV}=2.5$ at $\phi=0.54.$


To recap, under the reasonable assumptions that the wind is optically
thin and in temporary ionization equilibrium, we can estimate the
density by specifying the location of the wind ($n_{e}\propto
R^{-2}$). Based on the dynamical estimates of \citetalias{NL09}, we set 
$R_{\rm wind}\sim2.5\times10^{11}$ cm. We use photoionization
cross-sections from \citet{Verner96} and recombination rates from 
{\sc xstar} (T. Kallman, private communication). For temperatures
$10^{5}$ K $\le T\le10^{7}$ K, we find that the implied electron
density in units of cm$^{-3}$ is $10.8<\log(n_{\rm e})<12.4$. For the
measured luminosities we can expect the ionization parameter to be
between $10^{3}$ and $10^{5}.$ This is reasonable for a plasma
dominated by hydrogen- and helium-like iron.
\subsubsection{Ionization Parameter and Plasma Dynamics}
At the beginning of Section \ref{sec:diskwind}, we posed the question:
is it possible that the structure of the wind is constant in time, so
that the variability in the spectral lines is purely due to the
changing ionizing luminosity? Here we show that the answer is a
resounding `no'. This implies that the wind must be re-formed or
re-launched each and every cycle. Even without the photoionization
considerations that follow, this answer could be anticipated from
Figure \ref{fig:philine}, where it can be seen that between $\phi=0.2$
and $\phi=0.54,$ the flux in the Fe\,{\sc xxvi} and Fe\,{\sc xxv} lines
are roughly equal, while the X-ray luminosity increases by $\sim40\%$
over the same phase interval. It is quite unlikely that these observed
line fluxes could be produced purely by photoionization, because the
ion fractions for H- and He-like species are only comparable over
narrow intervals in $\xi$ \citep{Kallman01}.

That said, we performed a brief photoionization analysis with 
{\sc xstar}, using the analytic model {\sc warmabs}. Since we are
essentially modeling only H- and He-like iron, it should be understood 
that these models are designed not to give an exhaustive account of
the photoionization state of the gas but to parametrize the
relationship between the X-ray continuum and the wind. We use our
models of the ionizing spectrum (Section \ref{sec:pca}) at $\phi=0.54$
and $\phi=0.92$ to generate atomic level populations. We then fit our
6--7.5 keV HETGS spectra for the equivalent hydrogen column density,
ionization parameter, and velocity shift of the wind at those
phases. We fix the turbulent line width at $200$ km s$^{-1}$
(Section \ref{sec:hetgs}; see also \citealt{U09}). We set the density
at $n=10^{12}$ cm$^{-3}.$ Our best fits indicate that at $\phi=0.54,$
when Fe\,{\sc xxv} is maximized, the equivalent hydrogen column
density, ionization parameter, and velocity of the accretion disk wind
are $N_{\rm H}=2.1_{-0.8}^{+0.7}\times10^{22}$ cm$^{-2},
~\log\xi=3.87_{-0.02}^{+0.09}$ ergs cm s$^{-1},$ and
$v=-550_{-390}^{+200}$ km s$^{-1}$. The errors are $1\sigma.$ At
$\phi=0.92$, the maximum of Fe\,{\sc xxvi}, we find $N_{\rm H}=
10_{-2}^{+6}\times 10^{22}$ cm$^{-2},~\log\xi=4.8\pm0.1$ ergs cm
s$^{-1},$ and $v=-1520^{+230}_{-240}$ km s$^{-1}$ for this disk
wind. More detailed analysis would be necessary to account for
additional dynamical components at this phase
(Fig.\ \ref{fig:HETGS_spec}). 

We have already pointed out the changing velocity of the absorber, but
we notice here a significant increase in the column density and
ionization parameter of the wind (factors of roughly 5 and 7,
respectively). During this same phase interval ($\phi=0.54-0.92$) the
ionization parameter increases far more than the X-ray  
luminosity. The luminosity increases by a factor of 2, and the
ionizing flux for H- and He-like iron only increases by a factor
$\sim1.4-1.5.$  This is an indication that the luminosity variations are
insufficient to produce the observed absorption line variability. The
apparent changes in the fitted column density are further evidence that 
the structure of the wind cannot be constant in time. This, too, could
be anticipated from Figure \ref{fig:HETGS_spec}, where both the
wind ionization state and the strength of the Fe\,{\sc xxvi} line
change significantly with $\phi.$ 

Admittedly, the exact value of the column density could be
underestimated if resonance scattering or thermal emission lines fill
in some of the absorption features during the heartbeat cycle (see,
e.g.\ \citealt{Wojdowski03,Kallman09}). Additionally, the column of
Fe\,{\sc xxvi} may be enhanced by recombination in the totally ionized
wind. However, even if these effects are important, significant
ionization evolution in the disk wind is still implied by the fact
that Fe\,{\sc xxv} disappears while Fe\,{\sc xxvi} remains.
The appearance of a blue wing in the Fe\,{\sc xxvi} line profile near 
$\phi=0.92$ (Figure \ref{fig:HETGS_spec}) is a final indication that
the structure of the outflow varies substantially during the cycle,
probably due to the injection of additional material. 

\subsubsection{Wind Formation Scenarios}
\label{sec:formation}
The observed strong variability in the accretion disk wind leads to
two intriguing conclusions. First, the density and structure of the
wind must change on very short timescales. This must be the case
because the wind becomes progressively more ionized as the cycle
proceeds, but the X-ray luminosity does not increase enough to produce
the resulting ionization. Similarly, the distance between the wind and
the X-ray source cannot change much in the requisite $\sim20$ seconds
because the outflow is too slow. Thus equation (\ref{eq:ion}) dictates
that the density or structure of the wind must change. Second, these
changes in the wind must be quasi-periodic (as observed).
Fast quasi-periodic variability in the wind is rather unusual, so we
address it first. There are three basic methods by which the wind
might show coherent variability during the cycle:
\begin{enumerate}
\item Its structure is constant in time, but its ionization parameter
  changes in accordance with $L_{\rm X}.$ 
\item The wind is periodically launched from within or close to the
  oscillating region of the inner disk.
\item The wind is periodically launched from the outer disk (where
  previous studies have suggested it originates) by radiation from the
  inner disk. 
\end{enumerate}
Based on simple ionization arguments, we have already ruled out Case
1. Cases 2 and 3 are more complex scenarios, and we weigh their pros
and cons in what follows. We are unable to rigorously rule
out these scenarios, but the outer-disk origin seems preferable (see
below).

For Case 2, let us suppose the wind originates at $50~R_{\rm g}.$
Since the phase variations in this region are strong, an oscillating
wind seems natural. However, given our measured ionization parameters,
the density in the wind would have to be $n_{\rm e}\sim3\times(10^{17}-
10^{19})$ cm$^{-3}.$ Neither thermal nor radiation-driven winds can
produce such dense, highly ionized outflows at such small radii (see
\citet{PK02} and references therein), which would leave MHD processes
\citep{Proga2000,M06a,M08}. But even if the magnetic field does vary coherently
during the $\rho$ cycle, it is difficult to understand how a wind at
$50R_{\rm g}$ could have a velocity as low as 500 km s$^{-1}$ and a
line width $\lesssim 800$ km s$^{-1},$ given that the orbital velocity
at $50~ R_{\rm g}$ is approximately 40,000 km s$^{-1}$ for a 14
$M_{\sun}$ black hole with a spin parameter $a_{\star}=0.98$
\citep{McClintockShafee06}. It is also unclear how this wind could
persist for 30 ks given the short infall time at $50~R_{\rm g}.$ 

For Case 3, let us suppose that the wind originates at 10 lt-s from
the black hole. At this radius, dynamical and viscous timescales are
much too long to produce coherent phase variability. However, the
strong continuum variability provides a straightforward origin for the
wind: as the bright X-ray burst propagates outwards, it should
irradiate the disk in ever-larger annuli, launching a thermally driven
wind from each radius. This imprints a strong phase dependence on the
wind. In other words, the outer disk and inner disk are linked by
radiation.

The most significant challenge for this scenario involves
getting the wind into the line of sight quickly, since the inclination
of the disk (i.e.\ the jet axis) is 66$^{\circ}$. There are several
options to resolve this dilemma, all of which amount to reducing the
angle between the plane of the outer disk and the line of sight: (a)
Irradiated disks are known to flare vertically at large radii, with
scale heights $H/R\lesssim0.1;$ (b) radiation- or tidally-driven warps
may be present in the outer disk of GRS 1915+105
\citep{Pringle96,Whitehurst91}; (c) 
the inner disk and outer disk may be misaligned, as suggested by
\citet{L02} to connect the overabundance of iron along the line of
sight to the geometry of the supernova that formed the black hole
(e.g.\ \citealt{Brown00}). With a combination of these possibilities,
it may be possible to produce the observed 30--40 second rise time for
the wind. However, it should be noted that this scenario, while
favored, cannot be fully tested at this time because our HETGS phase
resolution is limited. We find that shifting our $\phi=0$ times by
some constant offset (on top of the random scatter reported in Section 
\ref{sec:period}) merely introduces a phase offset into Figure
\ref{fig:philine}. Longer observations of this wind may be able to
measure the travel time directly.

Case 3 also has the attractive potential to explain qualitatively a 
number of the observed properties of the disk wind, especially its
increasing column density, ionization parameter, and velocity. The
column density 
increase is simple: more material is launched into the light of sight
as time passes. In a thermally-driven wind, the velocity and
ionization parameter are proportional to
$R^{\gamma},$ and the density decreases like $n\propto R^{-2-\gamma}$
\citep*{B83}. $\gamma$ is a power-law index
between 0 and 0.5 under equilibrium conditions. If such scaling laws
are applicable in our observation, then the increase in $v$ and
$\xi$ may be partly related to thermal driving of the wind. A
robust estimate of $\gamma$ would require us to determine the precise
launch time of the wind, or the point at which the optical depth in
the newly-launched wind dominates that of the very highly-ionized wind
from the previous cycle. Such a probe is beyond the capabilities of
our present data. Simulations of the oscillation-driven wind (Proga et
al., in preparation) may shed additional light on these issues.

\subsubsection{The Role of the Wind in GRS 1915+105}
\label{sec:wind}
Finally, we attempt to estimate the mass-loss rate in the
wind in our observation. This is particularly important because recent
hydrodynamic simulations of thermally-driven winds \citep{Luketic10} have
shown that such winds can have mass loss rates well above the
accretion rate 
and thus may exert a strong influence on the accretion dynamics of the
system. Conditions here are likely to be far from steady state, so we 
use an elementary consideration based on the maximum column density in
the wind. In the cylindrical approximation, which is appropriate if
the wind is launched $\sim$vertically off the disk, $\dot 
M_{\rm wind}$ should scale approximately as
\begin{eqnarray}\label{eq:amnh}
\dot M_{\rm wind}&=&AmN_{\rm H}/\Delta t_{\rm launch}\\
&=&4\pi R_{\rm wind}v_{z}mN_{\rm H},\label{eq:mdotw}
\end{eqnarray}
where $\Delta t_{\rm launch}$ is the time over which the wind is
launched, $v_{z}$ is the wind speed perpendicular to the plane of the
disk, $A=2\times2\pi R_{\rm wind}v_{z}\Delta t_{\rm launch}$ is the 
outer surface area of the cylinder filled by the wind, and $m$ is the
average ion mass per hydrogen atom ($\sim2.4\times10^{-24}$ g assuming
ISM abundances from \citealt{Wilms00}). 

If $v_{z}\approx v\approx1000$ km s$^{-1}$ and $R_{\rm wind}=46$ lt-s,
then the implied wind mass loss rate is $\dot M_{\rm wind}\lesssim
3.9\times10^{20}$ g s$^{-1},$ which is roughly 25 times our
measured maximum mass accretion rate. In \citetalias{NL09}, we estimated $\dot
M_{\rm wind}$ based on spherical symmetry arguments and the ionization
parameter. Our estimate here is consistent with \citetalias{NL09} if the
covering factor in the spherical approximation is of order unity. This
may be more appropriate for the heartbeat state than for the other
\textit{Chandra} observations, since our average HETGS $\rho$-state
spectrum features a P-Cygni profile with roughly equal emission and
absorption components. The kinetic luminosity in the wind is $\dot
L_{\rm wind}\lesssim10^{36}$ ergs s$^{-1},$ which is far less than the
radiative luminosity. Still, if the wind is as massive as implied by
our estimates from equation (\ref{eq:mdotw}), we should expect it to
exert a significant influence on the accretion dynamics in GRS 1915+105.

\citet{Shields86} argued that very massive thermal winds can excite
long-period oscillations in the accretion disk as long as the mass
loss rate in the wind is sufficiently high ($\dot M_{\rm wind}
\gtrsim15 \dot M$) and increases with the accretion rate. As the
accretion rate rises, the wind becomes more massive and begins to
drain the disk. This emptying quenches the wind, allowing the disk to
refill, producing a repetitive cycle. Our $\dot M_{\rm wind}$ is
potentially high enough to produce this long instability, with a
period
\begin{equation}\label{eq:shields}
P\approx3400{\rm s}~\alpha^{-7/9}M^{14/9}T_{\rm IC8}^{-4/3}\dot M_{17}^{-1/3},
\end{equation}
where $M$ is the black hole mass in units of $M_{\sun},~T_{\rm IC8}$
is the Compton temperature in units of $10^{8}$ K, and $\dot M_{17}$
is the mass accretion rate in units of $10^{17}$ g s$^{-1}.$ 

Previously, X-ray \citep{Wilms01} and optical \citep*{Brocksopp01} observations
provided strong evidence that this instability may drive soft/hard
state transitions in the black hole binary LMC X-3, although no wind
was observed directly \citep{Cui02,Page03}. In GRS 1915+105, thermal winds
have been suggested as dynamically important (\citealt{L02};
\citealt*{Rau03}), but only 
recently has the contribution of this particular massive wind to X-ray
variability been measured \citep{NL09,Luketic10}. For our measured
parameters in the heartbeat state of GRS 1915+105, $P\sim1.5\times
10^{6}$ s, or about 17 days, if $T_{\rm IC8}=1$. To compare this
result to the available data on the $\rho$ oscillation, we locate all
observations of heartbeats in the entire \textit{RXTE} PCA archival
lightcurve of GRS 1915+105. Around each pointing, we search for nearby
$\rho$ states in windows of various durations. Statistically, we find
that 90\% of the pointed observations in a window of width $12\pm5$ days
surrounding a $\rho$ state observation also show heartbeats.
We conclude that the period of the Shields oscillation $P$ is
comparable to the typical occupation time for the $\rho$ state, so
that the wind-driven instability may actually be responsible for
transitions into/out of this state.

Our results therefore indicate that the accretion disk wind in GRS
1915+105 may play an integral role in transitions between variability
classes, effectively acting as a gatekeeper or a valve for the
external accretion rate, and facilitating or inhibiting state
transitions. In this context, it is notable that all published
observations of GRS 1915+105 in a remotely soft or variable state with
enough sensitivity to detect highly ionized absorption lines have done
so \citep{K00,L02,U09,NL09,U10}. We note that \citet{L02} actually
detected these lines in a relatively soft instance of the hard $\chi$
state, and that we showed in \citetalias{NL09} how the wind is sensitive to the
fractional hard X-ray flux, not the specific state.

Although the duration of our observation is clearly too short to probe 
long-term changes in the accretion flow, it provides us with a
useful diagnostic of the slower changes in the disk wind. Inspection
of a series of time-resolved spectra (subsets of our observation 3--10
ks in duration) indicates that the average ionization level of the
wind may decrease over the course of the 30 ks \textit{Chandra}
observation (not unlike the changes detected by
\citealt{L02,Schulz02,U09}). The origin of this change is unclear, and it
may have no immediate impact on state transitions: PCA pointed
observations show the source to be in the heartbeat state on 2001 May
23 (our observation) and May 30, but not May 16 or June 5. The
sampling is hardly sufficient to determine the time of state
transitions, but the ASM lightcurves suggest that this $\rho$ state
may have persisted for at least 3--5 days before and after the time
interval presented here. 

\begin{figure}
\centerline{\includegraphics[angle=270,width=0.49\textwidth]{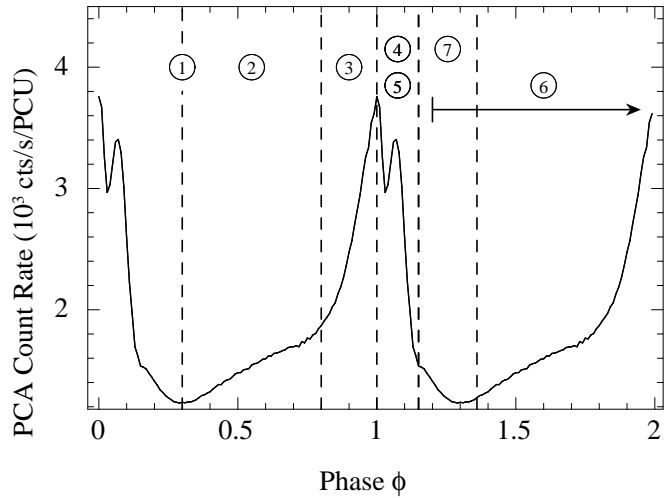}}
\caption{The PCA count rate lightcurve for the heartbeat state,
  labeled to correspond to our physical description of the $\rho$
  cycle in Section \ref{sec:physics}.}
\label{fig:summary}
\end{figure}
\section{TRACKING THE CYCLE PHYSICS}
\label{sec:physics}
\begin{deluxetable*}{lclccl}
\tabletypesize{\scriptsize}
\tablecaption{Physics of the $\rho$ Cycle\label{table:summary}}
\tablewidth{\textwidth}
\tablehead{
\colhead{Cycle}  & 
\colhead{Phase}  & 
\colhead{\S 6}  &
\colhead{Location} &
\colhead{Discussed}  &
\colhead{Observational}\\
\colhead{Event} & 
\colhead{Interval} & 
\colhead{Physics}  &
\colhead{($R/R_{\rm g}$)} &
\colhead{in \S {\sc x}}  &
\colhead{Evidence}}
\startdata
Minimum &$\sim$0.3 & 1: Density wave originates, & 25 & 
\ref{sec:rad} & Strong cyclic variations in $L_{\rm disk}$ and \\
\vspace{1mm}& & propagates in the inner disk && & accretion rate; High 
$L_{\rm disk}/L_{\rm Edd}$\\ 
Slow Rise &0.3--0.8 & 2: Local Eddington evolution &
1--5&\ref{sec:zsource}-\ref{sec:rad} & 
$L_{\rm disk},~R_{\rm in},$ and $\dot M$ grow slowly\\
\vspace{1mm}& &  &  && at roughly constant $T_{\rm obs}$\\
Soft Pulse &0.8--1.0 & 3: Disk luminosity rises &
1--5&\ref{sec:bremss}, \ref{sec:rad} &
\vspace{1mm}$\max(L_{\rm disk})\lesssim L_{\rm Edd}\tablenotemark{a}$\\ 
 &0.9--1.0 & 3: Disk becomes unstable,&1--5 & \ref{sec:bremss},& 
Unstable: $R_{\rm in}$ drops, $T_{\rm obs}$ spikes;\\
\vspace{-0.5mm} & & possibly due to radiation &&\ref{sec:zsource}-\ref{sec:rad} & Radiation Pressure: $L_{\rm disk}\lesssim L_{\rm Edd}\tablenotemark{a}$\\
\vspace{1mm}& & pressure& & & at constant high $\dot M$ ($L\propto T^{4/3})$\\
Hard Pulse &0.0--0.1 & 4: Disk ejects material, &1--5&
\ref{sec:bremss}, \ref{sec:eject} &
$E_{\rm fold},~kT_{\rm e}$ plummet; Bremsstrahlung\\
 & & which collides with corona& &  & normalization spikes\\
 &0.05--0.15 & 5: Density wave subsides, &1--5 & \ref{sec:mdot},& 
$L_{\rm disk}$ and $T_{\rm obs}$ drop, $R_{\rm in}$ grows \\
\vspace{1mm} & & disk relaxes &&\ref{sec:zsource}, \ref{sec:rad} & at
constant low $\dot M$ ($L\propto T^{4/3})$\\ 
Wind Formation &0.2--1.1 & 6: Intense X-ray heating &$10^{5}-10^{6}$&
\ref{sec:diskwind} & 
Fe\,{\sc xxv}, Fe\,{\sc xxvi} absorption lines \\
 & & launches, ionizes a wind & && grow and then fade; Line widths and \\
\vspace{1mm} & & from the outer disk && &  blueshifts are non-relativistic \\
Hard X-ray Tail &0.15--0.36 & 7: Production of a & $1-10^{3}$&
\ref{sec:powspec} & 
Band-limited noise, high $rms$; \\
\vspace{1mm} & & short-lived jet & &\ref{sec:model1}, \ref{sec:eject}
&hard X-ray spectrum; low luminosity
\enddata
\tablecomments{Physical processes in the $\rho$ cycle are arranged
  here by phase and by location in the accretion disk, along with
  observational evidence and relevant sections of the text. Quoted phase
  intervals are approximate.}
\tablenotetext{a}{Here $L_{\rm Edd}$ is the global Eddington limit,
  i.e.\ the Eddington limit for spherical accretion.\vspace{-1mm}}
\end{deluxetable*}
For clarity, we present a brief step-by-step summary of our model for
the physics of the heartbeat state, beginning at the minimum of the
cycle for ease of narration. Each step of this narrative is marked on
the X-ray lightcurve in Figure \ref{fig:summary}. Furthermore, each
event is included in Table \ref{table:summary} along with its relevant
phase interval, location in the text, and observational evidence. The
precise details apply specifically to this observation, but in the
future, it may be possible to generalize to other instances of this
oscillation.
\begin{enumerate}
\item ($\phi\sim0.3$) A wave of excess material, supplied by the high
  external accretion rate, originates near $25R_{\rm g}$ and
  propagates radially (inwards and outwards). For $R\lesssim100~ 
  R_{\rm g},$ the disk is always dominated by radiation pressure.
\item ($\phi=0.3-0.8$) Possibly as a result of local Eddington
  effects, the disk responds by increasing its inner radius (the slow
  rise) at $\sim$constant temperature. 
\item ($\phi=0.8-1.0$) During the soft pulse, the disk luminosity
  increases rapidly. When it reaches its maximum, $R_{\rm in}$ drops
  sharply, possibly at a constant accretion rate. This suggests a disk 
  instability at the end of the density wave. Incidentally, the
  maximum disk luminosity in this interval is $\sim90\%$ of the global
  Eddington limit for spherical accretion.
\item ($\phi=0.0-0.1$) The instability leads to the ejection of the
  material from the inner disk, possibly the hot surface layer, which
  may produce a flash of bremsstrahlung as it collides with the hot
  corona (the hard pulse).
\item ($\phi=0.05-0.15$) The wave of material subsides; the disk radius
  moves out quickly at the new (low) accretion rate.
\item ($\phi=0.2-1.1$) The bright X-ray pulses travel out along the
  disk, providing an impulse of X-ray heating and producing a hot,
  massive wind, which is subsequently over-ionized. Radiation thus
  links the dynamics of the inner and outer accretion disks from
  $R=1-10^{6}~R_{\rm g}$. 
\item ($\phi=0.15-0.36$) The hard X-ray spectrum, high $rms$, and
  band-limited noise near hard X-ray tail and the minimum of the cycle
  signal the possible production of a short-lived jet. Conditions in
  the accretion disk and corona are similar to other hard-state-like
  dips in GRS 1915+105.
\end{enumerate} 
\section{CONCLUSIONS}
\label{sec:conc}
In this paper, we have presented a detailed analysis of the spectral
and timing variability of GRS 1915+105 in the `heartbeat' ($\rho$)
state, a 50-second quasi-regular oscillation that usually consists of
single- or double-peaked bursts approaching the Eddington
luminosity (\citetalias{TCS97}, \citetalias{B00}). The accretion dynamics in this state are
particularly interesting because we find strong periodic variations in
both the X-ray continuum and in the accretion disk wind. Our results 
constitute the very first observational probe of accretion disk wind
physics on single-second timescales, nearly four orders of magnitude
shorter than the dynamical time at the location of the wind. By
performing our analysis with respect to cycle phase (as opposed to
time), we gain special insight into the physics that drives the
oscillation (the radiation pressure instability and the local
Eddington limit) and the origin of the accretion disk wind (transient
X-ray heating). We argue that our observation constitutes a snapshot
of an evolutionary process around this black hole, in which the wind
and its variability play an integral role in the changing state of the
accretion disk on timescales from seconds to weeks or months.

In the context of previous results, we have been able to answer two
pressing questions about X-ray variability in GRS 1915+105:
\begin{enumerate}
\item What is the origin of the hard pulse in the lightcurve of the
  $\rho$ state? Theoretical models predict single-peaked
  lightcurves (but see JC05), and yet single-peaked cycles
  account for only 40\% of the observed cycles (Neilsen et al.\ 2011,
  in preparation).
\item How does the wind know about the variability in the inner
  accretion disk? This question is extremely important given our 
  argument in \citetalias{NL09} that the wind functions as the
  mechanism for jet suppression.
\end{enumerate}

Our models of the X-ray continuum suggest that the hard pulse in
the lightcurve is produced by the radiation pressure-driven ejection
of material from the inner disk. As the ejected plasma collides with
the corona, it either produces a burst of bremsstrahlung or becomes so
Compton-thick that it scatters nearly all the light from the inner
disk. Simulations (NRM00, JC05) have predicted such ejections, but we
believe our result constitutes the first spectral detection of the
ejected plasma. This type of analysis may be repeated easily for many
other variability classes, allowing future detailed
characterizations of accretion/ejection physics by X-ray state. 

Of great interest is our ability to diagnose and understand these
X-ray states in the context of inflow and outflow physics. It is
particularly important to establish a clear link between the
variability of the inner accretion disk and the wind, without which a
connection between the wind and the jet would seem improbable. In our
analysis, we find that the fast variability in the X-ray luminosity of
the inner disk can actually produce significant structural changes in
the outer disk, producing a disk wind on the same timescales. Our
calculations based on equation (\ref{eq:shields}) provide an estimate of
(twice) the typical timescale for jet suppression if the wind actively
quenches the jet from the outer disk, but it is clear that both fast
and slow variations in the accretion rate cannot be neglected in this
process. Future observation of states with significant jet and wind
activity, like the $\beta$ state (\citealt{M98,NL09}; Neilsen et
al.\ 2011, in preparation), will reveal further details about the
precise nature of the wind-jet interaction.

Finally, we note that instruments like the High Timing Resolution
Spectrometer planned 
for the \textit{International X-ray Observatory (IXO)} will
revolutionize fast variability studies of both emission lines
(e.g.\ \citealt{Miller05}) and accretion disk winds in X-ray binaries. With
high signal-to-noise, good spectral resolution, and excellent time
resolution, it may be possible to perform QPO-phase-resolved
reverberation mapping with disk emission lines and wind absorption
lines, not only in GRS 1915+105, but in a large subset of the known
Galactic black holes. Furthermore, recent \textit{XMM-Newton} studies
of Seyfert galaxies have detected variable absorbers during long or
quasi-periodic dips (NGC 1365, \citealt{Risaliti09}; RE J1034+396,
\citealt{Maitra10}). The variations, which have been attributed to the
orbital motion of the warm absorber, suggest the exciting possibility
of a similar database of variability for supermassive black hole
systems. Future high-resolution spectral variability studies with
\textit{Chandra} and missions like \textit{IXO} will thus allow us to
probe even deeper into fundamental accretion physics on all mass
scales. 

\acknowledgements We thank the referee for comments which improved the
quality and clarity of our paper. We thank Claude Canizares, Norbert
Schulz, Tim Kallman, and Mike Nowak for helpful discussions of
spectral variability, and we thank Guy Pooley for making the radio
observations publicly available. J.N.\  and J.C.L.\ gratefully
acknowledge funding support from \textit{Chandra} grant AR0-11004X,
the Harvard University Graduate School of Arts and Sciences, and the
Harvard University Faculty of Arts and Sciences; R.A.R.\ acknowledges
partial support from the NASA contract to MIT for support of
\textit{RXTE} instruments.

{\it Facilities:} \facility{CXO(HETGS)}, \facility{RXTE(PCA)}
\appendix
\section{MEASURING PEAK TIMES}
\subsection{Motivation and Method}
Our primary concern in measuring the peak times for the $\rho$ cycle
is consistency between \textit{RXTE} and \textit{Chandra}. For RXTE
analysis alone, the single-peaked spectral hardness ratios may provide
a simpler method for measuring peak times. But since
\textit{Chandra}'s limited bandpass does not allow optimal hardness
ratios, we measure peak times from the X-ray count rate. As discussed
in Section \ref{sec:philc}, we use a representative cycle as a
template for cross-correlation with the entire lightcurve; the peaks
in the resulting cross-correlation  correspond to the peaks in the
lightcurve. We smooth the cross-correlation with a Gaussian of FWHM 3
seconds (much less than the typical width of a correlation peak) and
measure the time of maxima via parabolic interpolation. The resulting
maxima are preliminary peak times for the lightcurves. We use them to
create the average phase-folded lightcurve, which we then use as a new
template for cross-correlation. This process mitigates the possible
dependence of our results on our initial choice of template. 
\subsection{PCA versus \textit{Chandra}}
Given the high effective area of the PCA and the high flux of GRS
1915+105, we are confident in our peak times from \textit{RXTE}. 
However, because S/N in our grating spectra does not allow use to use
only the time intervals where both instruments are on-source, we need
to be equally confident in our peak times from \textit{Chandra}. We
actually find good quantitative agreement between the \textit{Chandra}
and PCA peak times during the overlap interval: the mean difference in
peak times is $\Delta T=-0.98\pm0.25$ s (where the error is the
1$\sigma$ sample standard deviation for the 273 peaks detected by both
instruments). Qualitatively, this means that the Chandra peak times
precede the \textit{RXTE} peak times and that the noise introduced
into the phase ephemeris by using the Chandra peak times, corrected by
$\Delta T,$ is much less than the cycle period (0.25 s $\ll$ 50 s). 
The fact that $\Delta T$ is non-zero reflects the different
instrumental sensitivities of the HETGS and the PCA, since the 2--6
keV lightcurves (normalized to mean count rate) have an RMS difference
of 0.1\%. The fact that $\Delta T\sim 1$ s is a numerical coincidence,
since the 2--4.5 keV PCA lightcurve peaks $1.47\pm0.19$ s before the
full PCA count rate. Physically, the fact that $\Delta T$ is negative
indicates that during this cycle, the hard flux peaks after the soft
flux. 
\bibliographystyle{apj}
\bibliography{ms}

\label{lastpage}

\end{document}